\pdfoutput=1
\documentclass[preprint,12pt]{elsarticle}

\usepackage{graphicx}
\usepackage{amssymb}
\usepackage{lineno}

\usepackage{times}
\usepackage{graphicx}
\usepackage[usenames]{xcolor}
\usepackage{amsmath}
\usepackage{amsfonts}
\usepackage{amssymb}
\providecommand{\U}[1]{\protect\rule{.1in}{.1in}}
\providecommand{\U}[1]{\protect\rule{.1in}{.1in}}

\journal{J. R. S. Interface}

\begin{document}

%\begin{frontmatter}


\title{The r\^{o}le of adhesion in contact mechanics}

\author{%%%% Author details
M.Ciavarella$^{1,3}$, J.Joe$^{2}$, A.Papangelo$^{1,3}$, and JR
Barber$^{2}$}

%%%%%%%%% Insert author address here
\address{
$^{1}$Polytechnic of Bari, Department of Mechanics, Mathematics and Management, Viale Japigia 182, 70126 Bari, Italy\\
$^{2}$University of Michigan, Department of Mechanical Engineering, Ann Arbor, MI 48109-2125, USA\\
$^{3}$Hamburg University of Technology, Department of Mechanical Engineering, Am Schwarzenberg-Campus 1, 21073 Hamburg, Germany}

%%%% Subject entries to be placed here %%
%%%% Keyword entries to be placed here %%%%
\begin{keyword}{Adhesion\sep contact mechanics\sep adhesion and fracture\sep rough contact\sep patterned surfaces}
\end{keyword}
%%%% Insert corresponding author and its email address}

%%%% Abstract text to be placed here %%%%%%%%%%%%
\begin{abstract}
Adhesive [e.g. van der Waals] forces were not generally taken into account in
contact mechanics until 1971, when Johnson, Kendall and Roberts (JKR)
generalized Hertz' solution for an elastic sphere using an energetic argument
which we now recognize to be analogous to that used in linear elastic fracture
mechanics. A significant result is that the load-displacement relation
exhibits instabilities in which approaching bodies `jump in' to contact,
whereas separated bodies `jump out' at a tensile `pull-off force'. The JKR
approach has since been widely used in other geometries, but at small length
scales or for stiffer materials it is found to be less accurate. In conformal
contact problems, other instabilities can occur, characterized by the
development of regular patterns of regions of large and small traction. All
these instabilities result in differences between loading and unloading curves
and consequent hysteretic energy losses. Adhesive contact mechanics has become increasingly important in recent years
with the focus on soft materials [which generally permit larger areas of the
interacting surfaces to come within the range of adhesive forces],
nano-devices and the analysis of bio-systems. Applications are found in
nature, such as insect attachment forces, in nano-manufacturing, and more
generally in industrial systems involving rubber or polymer contacts. In this
paper, we review the strengths and limitations of various methods for
analyzing contact problems involving adhesive tractions, with particular
reference to the effect of the inevitable roughness of the contacting surfaces.

\end{abstract}
%%%%%%%%%%%%%%%%%%%%%%%%%%%

%%%%%%%%%% Insert the texts which can accomdate on firstpage in the tag "fmtext" %%%%%

%\begin{fmtext}
%\end{fmtext}
%%%%%%%%%%%%%%% End of first page %%%%%%%%%%%%%%%%%%%%%

\maketitle


\section{Introduction}

\begin{figure}[th]
\centering
\includegraphics[height=50mm]{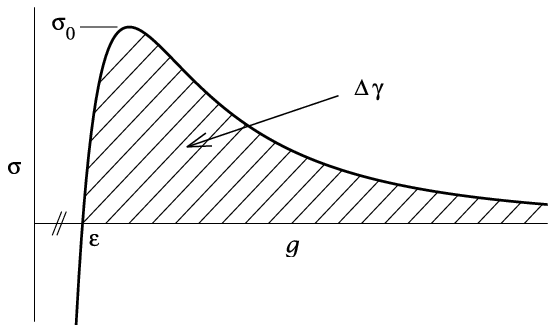}\caption{The Lennard-Jones traction law
between two half spaces. The interface energy $\Delta\gamma$ corresponds to
the shaded area.}%
\label{LJfigure}%
\end{figure}

Classical contact mechanics is typically characterized by the
\textit{Signorini inequalities}, which demand that the tractions between
interacting solid bodies be non-tensile, and that interpenetration of material
is inadmissible. We can then partition the surface of a body into regions of
\textit{contact}, where the gap between the bodies is zero and the normal
component of traction is compressive, and \textit{separation}, where there are
no tractions and the gap is positive. However, at very small length scales,
this dichotomy is an oversimplification. The local tractions between the
bodies will be a continuous function of relative approach and van der Waals
forces and other physical mechanisms can cause regions of tensile [or
\textit{adhesive}] tractions. Most authors assume that the Lennard-Jones 6--12
law \cite{LJ} defines the relation between the force and separation of two
individual molecules, and if a continuum is approximated as a uniform
distribution of molecules, the resulting traction $\sigma$ [tensile positive]
between two half spaces is then found to be \cite{Maugis}
\begin{equation}
\sigma(g)=\frac{8\Delta\gamma}{3\varepsilon}\left[  \frac{\varepsilon^{3}%
}{g^{3}}-\frac{\varepsilon^{9}}{g^{9}}\right]  \rule{5mm}{0mm}%
\mbox{where}\rule{5mm}{0mm}\Delta\gamma=\int_{\varepsilon}^{\infty}\sigma(g)dg
\label{LJ}%
\end{equation}
is the \textit{interface energy} or the work done per unit area of interface
in separating the two bodies from the equilibrium position $g=\varepsilon$, at
which $\sigma=0$. This expression is shown in Figure \ref{LJfigure}. The
maximum tensile traction occurs at a separation $g=3^{1/6}\varepsilon$ and is
$\sigma_{0}=16\Delta\gamma/9\sqrt{3}\varepsilon$.

\section{Contact of a sphere on a plane}

Equation (\ref{LJ}) can be integrated to determine the force transmitted
between two rigid bodies of known shape and relative position. Bradley
\cite{Bradley} used this method to determine the force between a rigid sphere
of radius $R$ and a rigid half plane. In particular, he showed that the
maximum tensile force [the \textit{pull-off force}] occurs when the point of
closest approach is equal to $\varepsilon$ and is of magnitude $2\pi
R\Delta\gamma$. This approach was extended by Rumpf \cite{Rumpf} and
Rabinovich \textit{et al.} \cite{Rabinovich} to estimate the adhesive force
between a small spherical particle and a rough surface, characterized as a set
of spherical asperities. These equations showed that even small amplitude of
roughness decreases pull-off by large factors.

\subsection{The JKR theory}

\label{SJKR}

\begin{figure}[th]
\centering
\includegraphics[height=60mm]{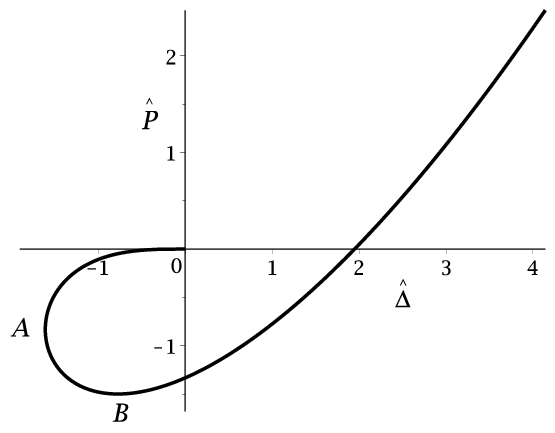}\caption{JKR solution for the relation
between dimensionless compressive force $\hat{P}$ and indentation $\hat
{\Delta}$ for the contact of a sphere and a plane.}%
\label{JKR}%
\end{figure}

If the contacting bodies are deformable, equation (\ref{LJ}) can be combined
with an analysis of the deformation, but the resulting boundary-value problem
is highly non-linear and generally can only be solved by numerical methods. An
approximation introduced by Johnson \textit{et al.} \cite{JKR} retains the
dichotomy between regions of contact and separation, but then computes the
total potential energy $\Pi=U+\Omega-\Gamma$ as the sum of elastic strain
energy $U$, potential energy of external forces $\Omega$ and interface energy
$\Gamma=A_{\mathrm{{c}}}\Delta\gamma$, where $A_{\mathrm{{c}}}$ is the total
contact area. The partition into areas of contact and separation is then
determined so as to minimize $\Pi$. This is now generally known as the
\textit{JKR solution}. Conceptually, it is identical to Griffith's theory of
fracture and hence is equivalent to linear elastic fracture mechanics [LEFM],
with $\Delta\gamma$ playing the r\^{o}le of the critical energy release rate
$G_{c}$. It follows that an alternative formulation is to demand that the
contact traction be square-root singular at all edges of the contact area,
with stress intensity factor
\begin{equation}
K_{\mathrm{{I}}}=\sqrt{2E^{\raisebox{0.7mm}{$ *$}}\Delta\gamma}\rule{5mm}{0mm}%
\mbox{where}\rule{5mm}{0mm}\frac{1}{E^{\raisebox{0.7mm}{$ *$}}}=\frac
{1-\nu_{1}^{2}}{E_{1}}+\frac{1-\nu_{2}^{2}}{E_{2}} \label{KI}%
\end{equation}
and $E_{i},\nu_{i}$ are Young's modulus and Poisson's ratio respectively for
the two bodies, with $i=1,2$. For the sphere, the relation between
[compressive] indentation force $P$ and indentation depth $\Delta$ can be
expressed for all cases in terms of the dimensionless parameters
\begin{equation}
\hat{P}=\frac{P}{\pi R\Delta\gamma}\;;\;\;\; \hat{\Delta}=\left(
\frac{E^{\raisebox{0.7mm}{$ *$}}R}{\Delta\gamma}\right)  ^{2/3}\frac{\Delta
}{R}\;, \label{eqJKR}%
\end{equation}
the resulting relation being shown in Fig. \ref{JKR}.

In this figure, the pull-off force under force control is defined by point $B$
and corresponds to $P=-3\pi R\Delta\gamma/2$, which differs from Bradley's
rigid-body value only by a factor $3/4$. Under force control, only points to
the right of $B$ are stable, whereas under displacement control, stability is
retained to the maximum negative indentation at $A$. In either case, once the
limiting point is reached, the sphere will jump out of contact and some energy
will be dissipated, presumably in the form of elastodynamic waves. Similarly,
if the sphere is slowly brought to approach the half space, it will jump into
contact from the origin to point $B$, again with a loss of energy. A sequence
of contact and separation cycles therefore implies a hysteretic loss of energy.

The original JKR solution considered only the contact of a sphere on a plane,
but the same technique can be applied to any geometry for which the
corresponding boundary-value problem can be solved. For example, Johnson
\cite{Johnson-Westergaard} gave the solution for a body with a sinusoidal
surface in partial contact with a plane. Also, the energetic argument can be
used to obtain numerical solutions using a boundary-element approach. For
example, Popov \textit{et al.} \cite{Popov-flat-adhesion} used this approach
to determine the pull-off force displacement relation for flat rigid punches
of various planforms. They showed that under displacement control, final
detachment occurs from a contact area approximately identified with a circle
inscribed in the planform, but that the maximum tensile force occurs before
this state is reached.

\subsection{A generalization of the JKR calculation}

\label{genJKR}

\begin{figure}[th]
\centering
\includegraphics[height=50mm]{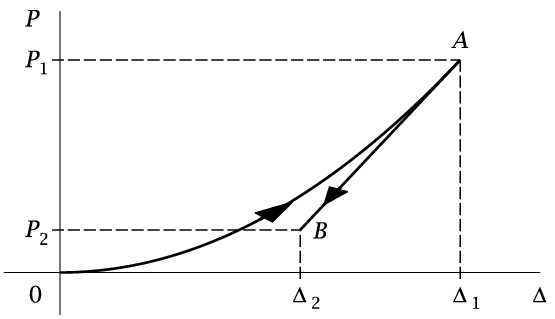} \caption{Two-step loading scenario.
(i) ``repulsive'' loading without adhesive forces until a given contact area
is reached (point A in the figure); (ii) Unloading at constant total contact
area up point B.}%
\label{JKRloading}%
\end{figure}

Johnson \textit{et al.} \cite{JKR} determined the elastic strain energy $U$
for the sphere problem by following the two-step scenario shown in Figure
\ref{JKRloading}. The contact is first loaded in compression to load $P_{1}$
establishing a contact area $A_{1}$. The contact area is then held constant
whilst the load is reduced to $P_{2}$. During this second phase, the load
displacement relation is linear as shown in the figure, and hence
\begin{equation}
P_{2}=P_{1}-(\Delta_{1}-\Delta_{2})\left(  \frac{\partial P}{\partial\Delta
}\right)  _{\Delta_{1}} \label{adhesive-load}%
\end{equation}

If the final value of displacement $\Delta_{2}$ is prescribed [displacement
control], the contact area $A_{1}$ and hence $\Delta_{1}$ must be chosen so as
to minimize the total potential energy
\begin{equation}
\Pi=U-A_{1}\Delta\gamma\
\end{equation}
We obtain
\begin{equation}
\frac{\partial\Pi}{\partial A_{1}}=0\rule{5mm}{0mm}%
\mbox{and hence}\rule{5mm}{0mm}\frac{\partial U}{\partial A_{1}}%
=\frac{\partial U}{\partial\Delta_{1}}\frac{\partial\Delta_{1}}{\partial
A_{1}}=\Delta\gamma
\end{equation}
Using Figure \ref{JKRloading} to determine $U$ as a function of $\Delta_{1},
\Delta_{2}$, we finally obtain
\begin{equation}
\Delta_{2}=\Delta_{1}-\sqrt{2\Delta\gamma\,\frac{\partial A_{1}}
{\partial\Delta_{1}}\left/  \frac{\mbox{$\partial$}^{2} P_{1}}
{\mbox{$\partial$} {\Delta_{1}}^{2}}\right.  }%
\end{equation}
\cite{CiaJKR2018}, which defines a general relation between the adhesive
solution and that without adhesion. Strictly, the argument requires that the
contact area be such as to give a uniform stress-intensity factor around the
perimeter as in axisymmetric problems, but it might reasonably be expected to
give good approximations in other cases.

\subsection{The Tabor parameter}

\label{STabor}

A numerical solution \cite{Muller} of the problem for a sphere using the
Lennard-Jones traction law of equation (\ref{LJ}) shows that the pull-off
force $P_{0}$ is a continuous function of the \textit{Tabor parameter}
\begin{equation}
\mu=\sqrt[3]{\frac{R(\Delta\gamma)^{2}}{E^{\raisebox{0.7mm}{$ *$}}%
\,^{2}\varepsilon^{3}}}\;, \label{mu}%
\end{equation}
\cite{Tabor}, tending to the Bradley rigid-body value of $2\pi R\Delta\gamma$
at $\mu=0$ and to the JKR value of $(3/2)\pi R\Delta\gamma$ as $\mu
\rightarrow\infty$. Equation (\ref{mu}) contains the radius $R$ and hence is
specific to the spherical contact problem. However, since the JKR solution is
formally identical to Linear Elastic Fracture Mechanics, a more general
expression can be obtained by analogy with the `small-scale yielding'
criterion. In particular, we can identify the width of the region in which the
predicted tensile traction exceeds the theoretical strength $\sigma_{0}$ as
\begin{equation}
s_{0}=\frac{E^{\raisebox{0.7mm}{$ *$}}\Delta\gamma}{\pi\sigma_{0}^{2}}\;.
\label{s0}%
\end{equation}
As in LEFM, the JKR solution is expected to provide a good approximation if
$s_{0}\ll a$, where $a$ is the smallest length scale associated with the
geometry of the problem --- e.g. the smallest width of the contact region, or
of the separation region. We can then define a generalized Tabor parameter
\begin{equation}
\mu=\sqrt{\frac{0.21a}{s_{0}}}\;, \label{gen-tabor}%
\end{equation}
where the numerical factor is included to ensure that it reduces to the
conventional definition in the case of the sphere, with $a$ then being the
radius of the circular contact area.

\subsection{Solutions for small $\mu$}

\label{SDMT}

When elastic effects are small, JKR solution is no longer appropriate and
there is a transition towards the rigid behaviour. Derjaguin \textit{et al.}
\cite{Derjaguin} gave an approximate solution of the spherical contact problem
by (i) finding the gap in the separation region for the classical Hertz problem
and then (ii) using the van der Waals [attractive] term from equation
(\ref{LJ}) to find the additional adhesive tractions and hence modify the
indenting force. This approach, known as the DMT solution, tends to the
rigid-body solution in the limit $\mu\rightarrow0$ and hence is often regarded
as an appropriate strategy when $\mu\ll1$. However, Pashley \cite{Pashley}
showed that the DMT approach can give unrealistic predictions [notably that
the pull-off force increases with $\mu$, and occurs at separations larger than
zero,\emph{ }which contradicts the results of rigorous numerical solutions],
and Greenwood \cite{Greenwood-DMT} offers an alternative approach based on
determining the elastic displacements due to the tractions predicted by the
rigid theory. We could also consider DMT as a first step of a more general
iterative strategy where the computed adhesive forces are allowed to generate
some deformation.

An alternative strategy is to approximate the traction law (\ref{LJ}) to make
the resulting boundary-value problem more tractable. Maugis \cite{Maugis-1992}
used a law in which the tractions are assumed to be constant and equal to the
maximum value $\sigma_{\max}$ from (\ref{LJ}) over a range $0<g<g_{\max}$,
beyond which they are zero. The value of $g_{\max}$ is chosen such that the
interface energy $\Delta\gamma=\sigma_{\max}g_{\max}$. This reduces the
contact problem to a linear three-part boundary-value problem which can be
solved in closed form for the case of the sphere. Alternatively, Greenwood and
Johnson \cite{double-Hertz} showed that the superposition of two axisymmetric
Hertzian traction distributions, one tensile and one [over a smaller circle]
compressive, could be chosen so as to satisfy the contact condition in the
smaller circle. The traction in the surrounding annulus is then a
single-valued function of gap and parameters can be chosen so as to ensure
that the maximum tensile traction is $\sigma_{\max}$ and the implied interface
energy is $\Delta\gamma$. Both these approaches predict a dependence of
pull-off force on $\mu$ qualitatively similar, but this may be due to the fact
that for the sphere, in the rigid limit there is no dependence on the form of
the force-separation law.\ Even for the sphere, the results are\emph{ }not
identical to the numerical solution \cite{Muller}.

\subsubsection{Bearing Area Method [BAM]}

\label{S-BAM}

A much simpler approximation \cite{BAM} appropriate for $\mu\ll1$ is to use
the Maugis-Dugdale force law \cite{Maugis-1992}, but to estimate the
attractive area $A_{\mathrm{{att}}}$ [i.e. the area in which there is
separation, but where $0<g<g_{\max}$] as
\begin{equation}
A_{\mathrm{{att}}}(\Delta)\approx B(\Delta+g_{\max})-B(\Delta)\;,
\label{BAM01}%
\end{equation}
where $B(\Delta)$ is the \textit{bearing area} --- i.e. the area over which
the bodies would need to interpenetrate each other if they were moved together
through a distance $\Delta$ and there were no elastic deformation. The total
compressive force applied to the indenter is then estimated as
\begin{equation}
P(\Delta)=P_{C}(\Delta)-\sigma_{\max}A_{\mathrm{{att}}}(\Delta)\;,
\label{BAM02}%
\end{equation}
where $P_{C}(\Delta)$ is the compressive force in the corresponding elastic
contact problem without adhesion. For the sphere, this procedure gives exactly
the same force-displacement relation as that denoted by DMT-M by Maugis
\cite{Maugis}.

\subsection{Effect of plastic deformation}

\label{S-plasticity}

If plastic deformation occurs during compressive loading, the effective
(unloaded) profile of the contacting bodies is modified and this affects the
pull-off force. If there is extensive plastic deformation, a crude
approximation can be obtained by assuming that contact pressure $p_{0}$ at
maximum compressive load $P_{0}$ is approximately uniform and equal to the
hardness $H$, so the contact area is a circle of radius $a_{0}$ (larger than
the Hertzian radius) where
\begin{equation}
P_{0}=\pi a_{0}^{2}H\;.
\end{equation}
We also note that $H\approx3\sigma_{Y}$, where $\sigma_{Y}$ is the uniaxial
yield stress. Johnson \cite{Johnson1976} used this result to estimate an
equivalent radius $R^{\prime}$ for the sphere after elastic unloading as
\begin{equation}
R^{\prime}=\frac{4E^{\ast}a_{0}}{3\pi H}\;,
\end{equation}
by assuming that unloading is approximately defined by the Hertzian analysis
from load $P_{0}$ and contact radius $a_{0}$. Using this value in the JKR
solution (\ref{eqJKR}) he then obtained an increased pull-off force
\begin{equation}
P_{c}^{\prime}=\frac{3\pi R^{\prime}\Delta\gamma}{2}=\frac
{2E^{\raisebox{0.7mm}{$ *$}}\Delta\gamma a_{0}}{H}=2E^{\raisebox{0.7mm}{$ *$}}%
\Delta\gamma\sqrt{\frac{P_{0}}{\pi H^{3}}}\;. \label{P'c}%
\end{equation}
Notice that the pull-off force now depends on the material properties
$H,E^{\raisebox{0.7mm}{$ *$}}$ and also increases with the square root of the
maximum load during initial compression.

A more precise solution requires a full analysis of the elastic-plastic
loading process, in particular to determine the exact contact radius and
traction distribution at maximum compressive load, followed by the exact
elastic solution corresponding to unloading from this condition. Mesarovic and
Johnson \cite{Mesarovic} showed that even at large compressive preloads,
Johnson's approximation (\ref{P'c}) underestimates the JKR pull-off force by a
factor of $3\pi/4$ due to Johnson's unrealistic Hertzian assumption during unloading.

They also presented results using the Maugis-Dugdale cohesive-zone model,
which they characterized in a space defined by the dimensionless parameters
\begin{equation}
\chi=\frac{\pi}{(2\pi-4)}\frac{E^{\raisebox{0.7mm}{$ *$}}\Delta\gamma}%
{p_{0}^{2}a_{0}}\;;\;\;\;S=\frac{\sigma_{0}}{p_{0}}\;.
\end{equation}
The pull-off force is always close to the JKR value, to which it tends
asymptotically when $\chi^{2/3}/S^{2}\rightarrow0$.

\section{Thin elastic layers}

Many engineering and scientific applications involve thin deformable layers
supported by a relatively rigid foundation. Examples include rubber layers
bonded to steel components and cartilage layers attached to bones. If an
elastic layer of thickness $h$ is bonded to a rigid foundation and then
subjected to a uniform tensile traction $\sigma$, the only non-zero strain
will be that in the thickness direction and the surface will move outwards
through a distance
\begin{equation}
u=\frac{\sigma}{k}\rule{5mm}{0mm}\mbox{where}k=\frac{E(1-\nu)}{(1+\nu
)(1-2\nu)h}\;. \label{k}%
\end{equation}
Johnson \cite{Johnson} argued that this remains a good approximation under
more general spatially-varying tractions as long as the layer is `sufficiently
thin' meaning that $h$ is small compared with the linear dimensions of the
loaded area. The layer then acts like a Winkler foundation of `modulus' $k$,
with proportionality between local displacement and local traction. In
particular, in non-conformal contact problems [such as indentation by a
sphere], the contact pressure then goes to zero at the edge of the contact area.

Johnson's argument was extended to problems involving adhesive tractions by
Yang \cite{Yang-comp} and Argatov \textit{et al.} \cite{Argatov}, using an
energy argument analogous to that in the JKR theory \cite{JKR}. They showed
that the effect of interface energy was to change the boundary condition at
the edge of the contact area from $\sigma=0$ to $\sigma=\sqrt{2k\Delta\gamma}%
$, which is independent of the contact geometry [as is the stress intensity
factor (\ref{KI}) in the JKR theory].

If $\nu\rightarrow0.5$, the modulus $k\rightarrow\infty$, since the layer
becomes incompressible. Deformation is still possible under non-uniform
tractions, but involves the displacement of material in the plane of the layer
\cite{Johnson}. Approximate solutions for the case with adhesive tractions are
given by Yang \cite{Yang-incomp}, Argatov \textit{et al.} \cite{Argatov} and
Papangelo \cite{Papangelo2018}.

\subsection{Instabilities}

\label{uniform}

The Lennard-Jones traction law $\sigma(g)$ of equation (\ref{LJ}) and Figure
\ref{LJfigure} can be regarded as a non-linear spring [with ranges of negative
stiffness] in series with the linear spring associated with the modulus $k$ of
equation (\ref{k}). For example, if $\Delta$ denotes the gap that would exist
between a layer and a plane surface in the absence of elastic deformation, the
actual gap will be $g$ where
\begin{equation}
\Delta=g+\frac{\sigma(g)}{k}\rule{5mm}{0mm}\mbox{and hence}\rule{5mm}{0mm}%
\frac{\partial\Delta}{\partial g}=1+\frac{1}{k}\frac{\partial\sigma}{\partial
g}\;. \label{Delta}%
\end{equation}
The gap $g$ and hence the traction $\sigma(g)$ will be multivalued functions
of rigid-body approach $\Delta$ if there exist ranges where
$\mbox{$\partial$}\Delta/\mbox{$\partial$} g<0$. Notice [for example from
Figure \ref{LJfigure}] that any traction law involving adhesive tractions must
exhibit a range of values of $g$ in which the slope $\mbox{$\partial$}\sigma
/\mbox{$\partial$} g<0$, and from (\ref{Delta}), instability is most likely to
occur at the point where the magnitude of this negative slope is maximum. For
the Lennard-Jones traction law of equation (\ref{LJ}), this maximum occurs at
$g=(15/2)^{1/6}\varepsilon$ and is of magnitude $1.253\Delta\gamma
/\varepsilon^{2}$.

\begin{figure}[th]
\centering
\vspace{3mm} \includegraphics[height=55mm]{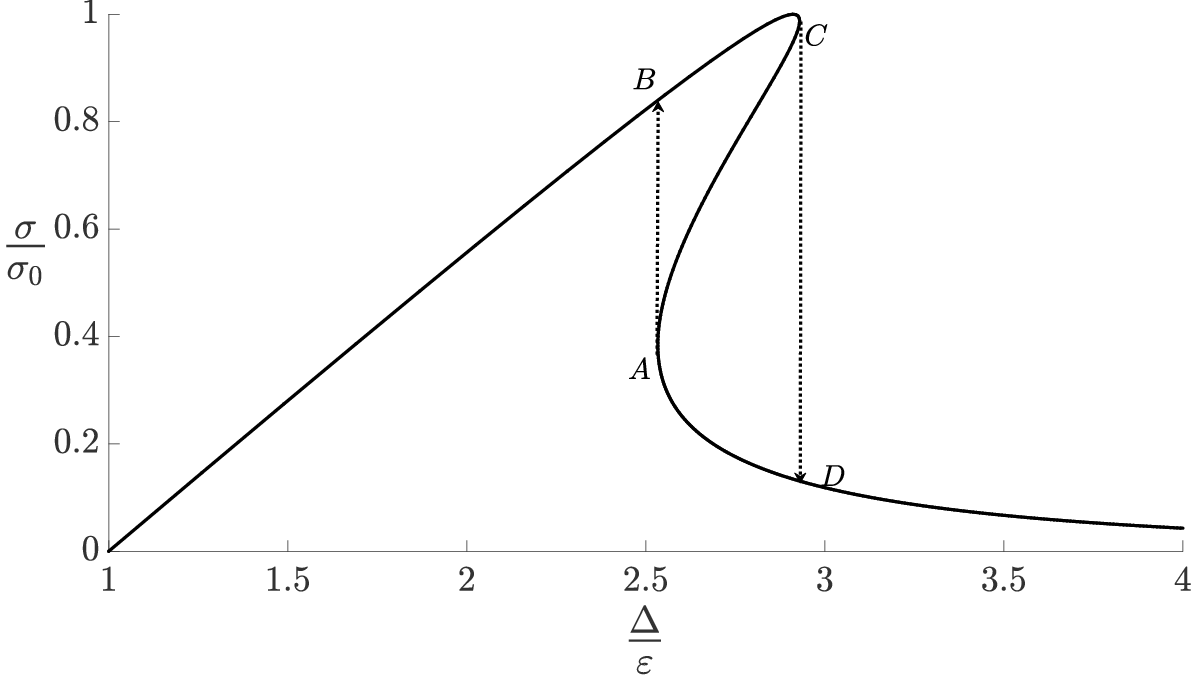} \caption{Adhesive
traction $\sigma$ as a function of rigid-body approach $\Delta$ for a layer
with $\nu=0.25$ and $\beta=0.5$. Jumps occur during approach from $A$ to $B$
and during separation from $C$ to $D$ as indicated by the dotted lines.}%
\label{jump-in-and-out}%
\end{figure}

A typical case involving instability is illustrated in Figure
\ref{jump-in-and-out}, where $\nu=0.25$ and the dimensionless parameter
\begin{equation}
\beta=\frac{E\varepsilon^{2}}{h\Delta\gamma} \label{beta}%
\end{equation}
is equal to 0.5. If the bodies are initially widely separated, the tractions
will be defined by the lower branch of the curve, but if $\Delta$ is reduced
below $\Delta_{A}$, there must then be a jump to the point $B$. A jump in the
opposite direction from $C$ to $D$ is anticipated during subsequent
separation, so that during an approach-separation cycle, there will be a
hysteretic energy loss defined by the area $ABCD$.

\subsection{Sinusoidal instabilities}

\label{S-instabilities}

The uniform state defined by equation (\ref{Delta}) can be unstable to
non-uniform perturbations even where jumps are not predicted. It is convenient
to define a dimensionless coordinate $\xi=x/h$ in the plane of the layer. For
a linear elastic layer, a sinusoidal traction distribution $\sigma(\xi
)=S\cos(\zeta\xi)$ will produce a surface displacement
\begin{equation}
u(\xi)=\frac{S}{k(\zeta)}\cos(\zeta\xi)\;,
\end{equation}
where $k(\zeta)$ is a wavenumber-dependent stiffness. For a uniform elastic
layer bonded to a rigid foundation, we have
\begin{equation}
k(\zeta)=\frac{E\zeta\left[  (3-4\nu)\cosh(2\zeta)+2\zeta^{2}+5-12\nu+8\nu
^{2}\right]  }{2h(1-\nu^{2})\left[  (3-4\nu)\sinh(2\zeta)-2\zeta\right]  }
\label{k(zeta)}%
\end{equation}
\cite{Hannah}, which reduces to (\ref{k}) in the limit $\zeta\rightarrow0$.
However, similar arguments can also be applied to more complex elastic systems
such as multilayers or functionally graded layers, the only change being in
the function $k(\zeta)$.

If the layer is placed such that the uniform solution of Section \ref{uniform}
involves a gap $g_{1}$, energetic arguments can be used to show that
infinitesimal sinusoidal perturbations on this solution of wavenumber $\zeta$
will then be unstable if
\begin{equation}
\left(  \frac{\partial\sigma}{\partial g}\right)  _{g=g_{1}}<-k(\zeta)\;.
\label{stability-inequality}%
\end{equation}
With the Lennard-Jones traction law, the uniform traction state will be
unconditionally stable provided
\begin{equation}
k(\zeta)>\frac{1.253\Delta\gamma}{\varepsilon^{2}}\rule{5mm}{0mm}%
\mbox{or equivalently}\rule{5mm}{0mm}\frac{E}{hk(\zeta)}<1.253\beta
\end{equation}
for all $\zeta$, from equations (\ref{beta}, \ref{stability-inequality}).

Non-linearity of the traction law (\ref{LJ}) places limits on the growth of
such a perturbation, but if the condition (\ref{stability-inequality}) is
satisfied for some range of values of $g_{1},\zeta$, we might then anticipate
the development of a spatially-periodic deformation pattern during approach of
the layer to a plane surface.

Patterns of this kind have been predicted theoretically
\cite{shenoy2001pattern,sarkar2004patterns} and observed experimentally
\cite{gonuguntla2006control,monch2001elastic} mainly for incompressible layers
for which the `uniform' instability of Section \ref{uniform} is suppressed.
Gonuguntla \textit{et al.} \cite{gonuguntla2006control} have shown how this
self-patterning behaviour can be used in the manufacture of patterned layers
using lithography.

\subsection{Periodic deformation patterns}

\label{S-pattern}

\begin{figure}[!th]
\hspace{-15mm}\includegraphics[height=75mm]{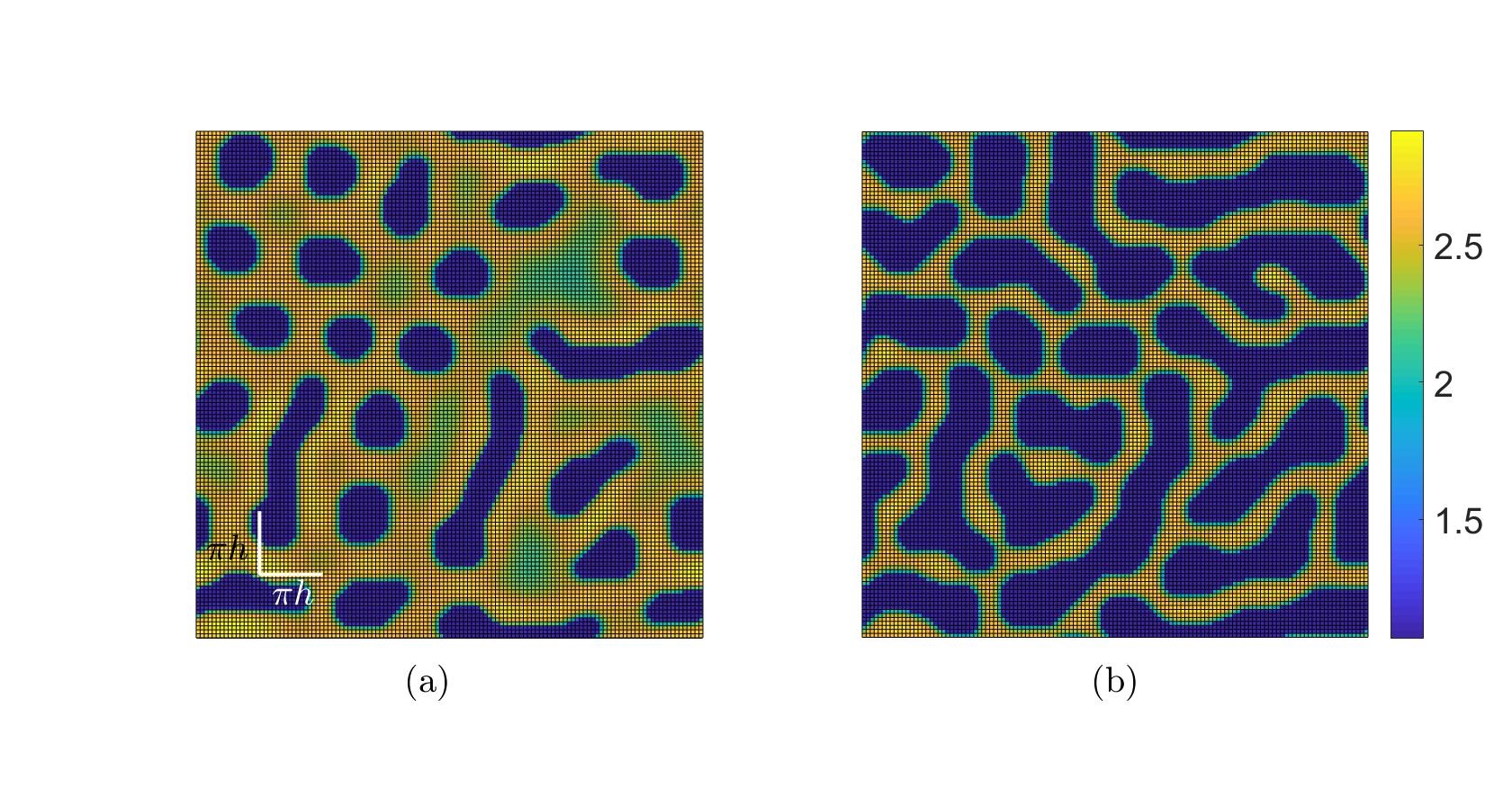}
\par
\vspace{-8mm} \hspace{-15mm}\includegraphics[height=75mm]{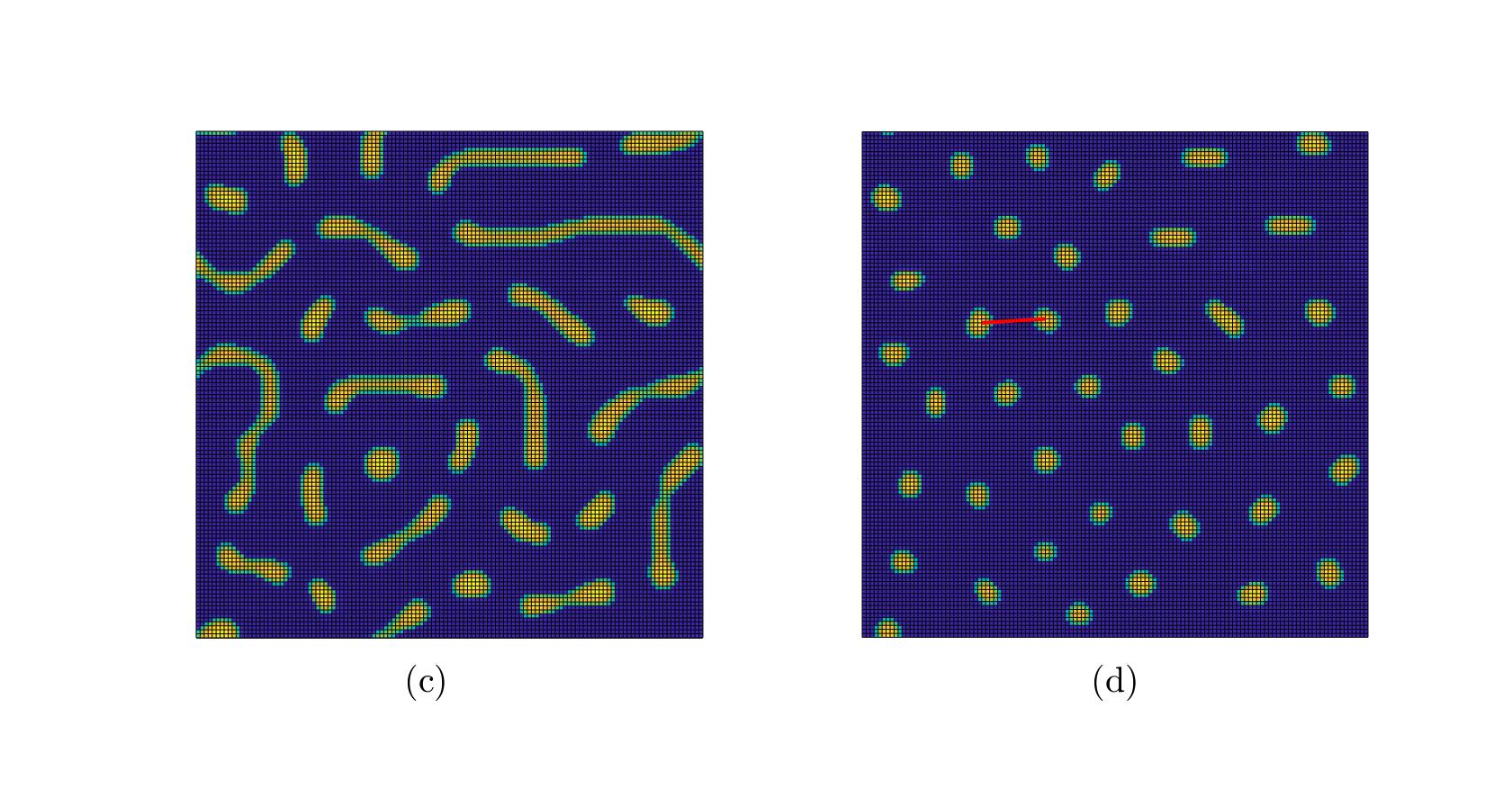}
\par
\vspace{-5mm}\caption{Contours of dimensionless local gap $g(x,y)/\varepsilon$
during approach of a uniform incompressible layer to a plane surface.
$\Delta=2.1\varepsilon$ (a), $1.7\varepsilon$ (b), $1.4\varepsilon$ (c),
$1.2\varepsilon$ (d). The contour scale applies to all four figures.}%
\label{pattern}%
\end{figure}

Figure \ref{pattern} shows contours of the gap $g(x,y)$ for four stages of
approach $\Delta$ for a layer with $\beta=5, \nu=0.5$. These results were
obtained using the Green's Function Molecular Dynamics [GFMD] algorithm of
Persson and Scaraggi \cite{Scaraggi}. Since the material is incompressible,
uniform instabilities of the type discussed in Section 3.1 cannot occur. The
contours are defined as multiples of $\varepsilon$ and the scale bar in Figure
\ref{pattern}(a) represents the layer thickness $h$.

During approach [$\Delta$ decreasing] the morphology is first defined by
pillars of 'contact' [values of $g$ close to $\varepsilon$] surrounded by
regions of much larger gap (a). Further reduction in $\Delta$ leads to the
labyrinth pattern (b) and then an inverted labyrinth (c) where regions of
contact are connected. The last stage (d) comprises a pattern of approximately
circular separation regions surrounded by contact. The red line in Figure
\ref{pattern}(d) represents wavelength corresponding to the most unstable
sinusoidal perturbation [see Section \ref{S35} and Figure \ref{k-zeta} below].

\begin{figure}[th]
\vspace{3mm} \includegraphics[height=60mm]{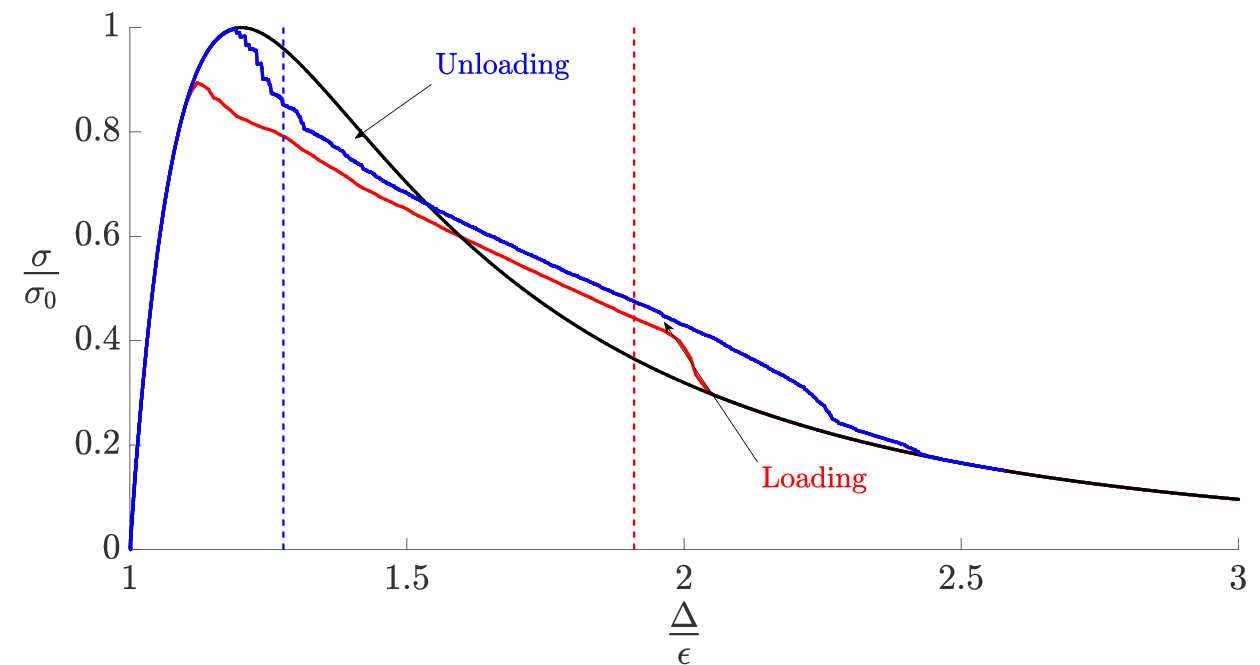}\caption{Relation
between mean traction and rigid-body separation $\Delta$ for $\beta=5,\nu
=0.5$. The uniform solution is unstable between the vertical dashed lines.}%
\label{hysteresis}%
\end{figure}

Theoretically, instability starts at the value of $\Delta$ at which an
infinitesimal sinusoidal perturbation first becomes unstable. However, once a
pattern is established, it persists beyond the range of linear instability and
hence the traction curves for loading and unloading are different, as shown in
Figure \ref{hysteresis}. During progress from contact to separation, patterns
develop before the theoretical point, presumably due to the use of finite
increments in the iterative algorithm.

\subsection{Determination of patterns using series methods}

An alternative approach for approximating these patterns is to represent the
elastic deformation as a finite Fourier series and use the Rayleigh-Ritz
method for determining the coefficients. For example, in two dimensions we
write
\begin{equation}
u(\xi)=\sum_{n=0}^{N} u_{n}\cos(n\zeta_{0}\xi)\;,
\end{equation}
where $\zeta_{0}$ is a fundamental wavenumber that might be related to the
finite dimension of the contact surface. The elastic strain energy per unit
area is then
\begin{equation}
U=\frac{1}{2}k(0)u_{0}^{2}+\frac{1}{4}\sum_{n=1}^{N} k(n\zeta_{0})u_{n}^{2}\;.
\end{equation}
and the interface energy per unit area is defined by
\begin{equation}
\Gamma=\frac{1}{2\pi}\int_{0}^{2\pi/\zeta_{0}}d\xi\int_{\Delta-u(\xi)}%
^{\infty}\sigma(g)dg\;.
\end{equation}
The coefficients $u_{n}$ are then determined using an appropriate optimization
algorithm so as to minimize the total energy $\Pi=U-\Gamma$ for a given value
of approach $\Delta$ \cite{sarkar2004patterns}. The same technique was
extended to three-dimensional patterns by Gonuguntla \textit{et al.}
\cite{gonuguntla2006control} using a double Fourier series.

\subsection{Effect of material parameters}

\label{S35}

\begin{figure}[th]
\centering
\vspace{3mm} \includegraphics[height=60mm]{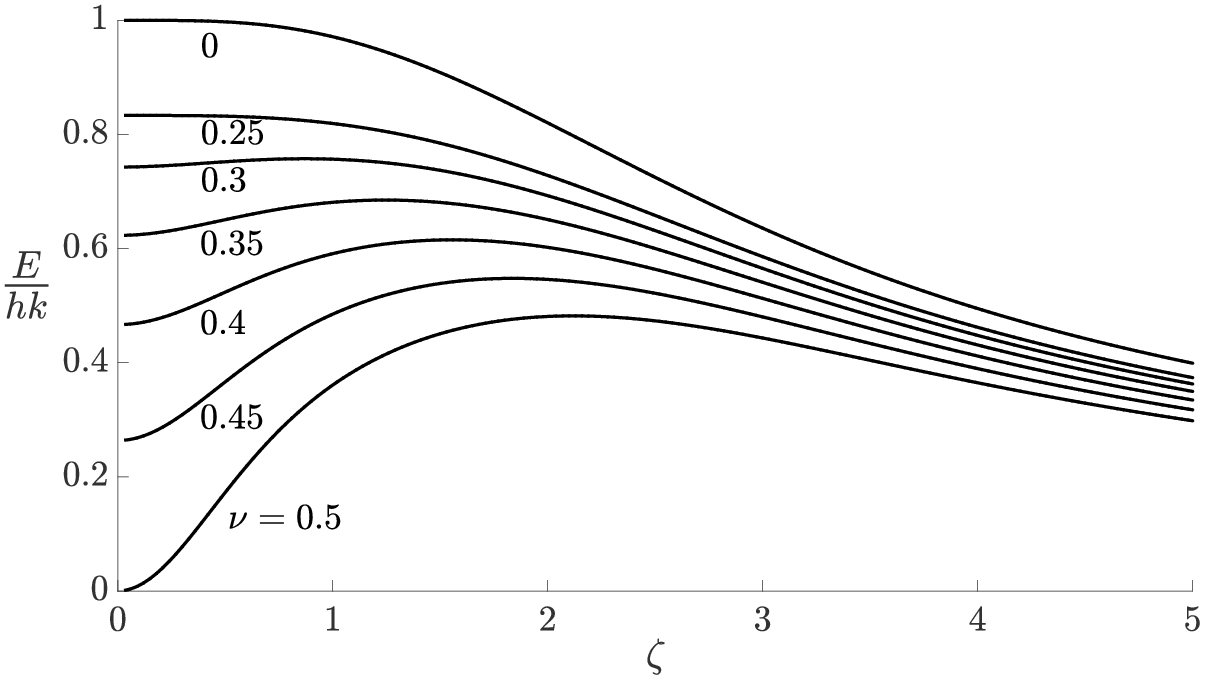}\caption{Dimensionless
layer compliance as a function of wavenumber for an elastic layer bonded to a
rigid foundation.}%
\label{k-zeta}%
\end{figure}

Figure \ref{k-zeta} shows the dimensionless layer compliance [reciprocal of
stiffness] $E/hk(\zeta)$ as a function of wavenumber $\zeta$ for the bonded
layer defined by equation (\ref{k(zeta)}) for various values of Poisson's
ratio $\nu$.

We notice from Figure \ref{k-zeta} that the curves for $\nu>0.25$ exhibit a
maximum at some value $\zeta=\zeta_{0}>0$, whereas for $\nu\leq0.25$, the
maximum occurs at $\zeta=0$. In both cases, instability will commence when the
gap $g=g_{1}$ in the uniform solution reaches the value at which
$\mbox{$\partial$}\sigma/\mbox{$\partial$} g$ first satisfies
(\ref{stability-inequality}). For $\nu\leq0.25$ the first unstable condition
corresponds to a uniform perturbation [$\zeta=0$] and hence occurs at the
point $A$ during loading and $C$ during unloading in Figure
\ref{jump-in-and-out}. The unstable response comprises a sudden change [jump]
in uniform traction as indicated.

\begin{figure}[th]
\centering
%\vspace{30mm}
\includegraphics[height=60mm]{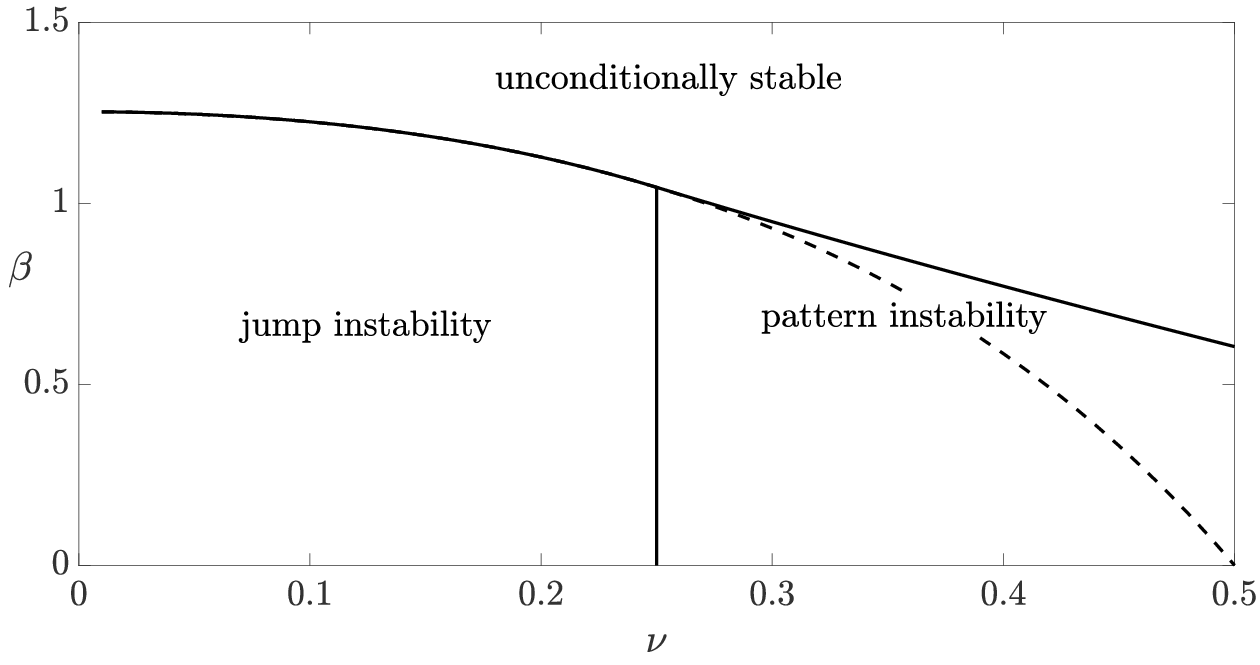} \caption{Dependence of stability
behaviour on $\nu$ and $\beta$.}%
\label{nu-beta}%
\end{figure}

For $\nu>0.25$, the maximum compliance occurs at a non-zero wavenumber
$\zeta_{0}$ and we anticipate the development of a pattern with this
periodicity, at least near the value of $g$ at which
(\ref{stability-inequality}) is first satisfied. This behaviour is shown
schematically in Figure \ref{nu-beta} for the Lennard-Jones traction law.

The dashed line in this figure defines the value of $\beta$ below which a
uniform perturbation [$\zeta=0$] is also unstable. In this region, the
non-uniform instability is triggered before the uniform one and generally
dominates the subsequent behaviour. However, this requires that a
representative in-plane dimension $L$ of the layer be large enough to
accommodate at least one wavelength of an unstable sinusoidal perturbation. In
most practical cases $h\ll L$ and which ensures that this condition is
satisfied except for values of $\nu$ quite close to 0.5.

Similar calculations can be performed for more complex layers. In particular,
we note that for a bi-material layer, the dimensionless compliance may exhibit
two distinct maxima \cite{bi-layer}. In such cases, the absolute maximum of
the curve defines the first instability during either approach or separation
and generally dominates the subsequent pattern development.

\section{Effect of roughness on adhesion}

If surfaces were perfectly plane, van der Waals forces would imply that two
such bodies brought into contact would be indistinguishable from a single
monolithic body and could only be separated by a fracture process involving
the application of tractions equal to the theoretical strength $\sigma_{0}$.
The fact that this doesn't generally happen is due to the microscopic
roughness of practical surfaces. Numerous authors \cite{Vakis} have developed
models to characterize and quantify the effect of surface roughness on solid
contact, often motivated by the attempt to explain Amontons' law of friction.
However, most of these models are based on the Signorini dichotomy between
contact and separation, and do not include adhesive [tensile] tractions. For
`stiff' materials such as metals, with macroscopic roughness (typically much
larger in amplitude than the range of attractive tractions), this assumption
is reasonable, but recent emphasis on flexible materials such as polymers and
biotissues makes the interaction between adhesion and roughness particularly relevant.

\subsection{Fuller and Tabor}

The first theoretical investigation of the effect of roughness on adhesion was
that of Fuller and Tabor \cite{FullerTabor}, who followed Greenwood and
Williamson [GW] \cite{Greenwood1966} in modelling the rough surface as a set
of identical spherical asperities of radius $R$ whose peak heights follow a
Gaussian distribution with standard deviation $h_{\mathrm{{rms}}}$, but who
used the JKR solution of Section 2.1 to describe the individual asperity
contacts. They predicted that the pull-off force should decrease drastically
with $h_{\mathrm{{rms}}}$ and that this effect is characterized by the
dimensionless parameter
\begin{equation}
\theta_{FT}=\frac{h_{\mathrm{{rms}}}^{3/2}\Delta\gamma}{R^{1/2}%
E^{\raisebox{0.7mm}{$ *$}}} \label{FT}%
\end{equation}
where we note that $\Delta\gamma/E^{\raisebox{0.7mm}{$ *$}}$ defines a
characteristic adhesion length which for contact of similar materials is
related to $\varepsilon$ of equation (\ref{LJ}).

\begin{figure}[th]
\centering
%\vspace{30mm}
\includegraphics[height=65mm]{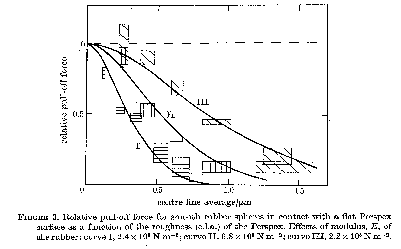} \caption{Relative pull-off decay
force for three different elastic moduli and increasing center line average
roughness in Fuller and Tabor experiments (from \cite{FullerTabor})}%
\label{FigureFT1975}%
\end{figure}

Fuller and Tabor also conducted experiments with rubber spheres contacting a
perspex plane with varying roughness amplitudes and obtained results that
correlate well with the theory (see Fig. \ref{FigureFT1975}). This is
surprising in view of the fact that the Fuller and Tabor theory suffers from
even more limitations than the original GW theory \cite{Greenwood2017} and the
theory was developed for the contact of nominally plane bodies, rather than
for a sphere on a plane. Strictly speaking therefore, the good qualitative
experimental agreement might be considered fortuitous.

\subsection{Fractal surfaces}

\begin{figure}[th]
\centering
\includegraphics[height=60mm]{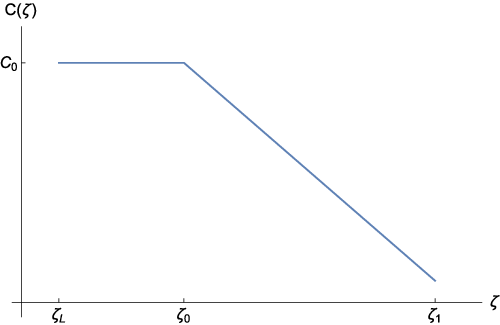} \caption{Log-log plot of the typical
spectrum of surface roughness which is used today to model idealized nominally
flat Gaussian rough fractal surfaces.}%
\label{PSD}%
\end{figure}

A not obvious aspect of Fuller and Tabor's adhesion parameter (\ref{FT}) is
that it contains the asperity radius $R$, which for random rough surfaces is
proportional to $m_{4}^{-1/2}$ \cite{McCool}, where $m_{4}$ is the fourth
moment of the Power Spectral Density (PSD) of the profile. However, modern
studies of nominally flat rough surfaces often assume a Gaussian isotropic 2D
roughness with a PSD of the power-law form
\begin{equation}
C\left(  \zeta\right)  =\left\{
\begin{array}
[c]{lc}%
C_{0}, & \zeta_{L}<\zeta<\zeta_{0}\\
C_{0}\left(  \frac{\zeta}{\zeta_{0}}\right)  ^{-2\left(  H+1\right)  }, &
\zeta_{0}<\zeta<\zeta_{1}\\
0\;, & \zeta>\zeta_{1}%
\end{array}
\right.  \label{eqPSD}%
\end{equation}
as shown in Fig. \ref{PSD}, where $\zeta$ is the wavenumber and to make the
surface more closely Gaussian, we have introduced a `roll-off' region of
constant PSD. The power-law segment in equation (\ref{eqPSD}) typically
extends over three or four decades so the roughness has a profound
`multiscale' character. The slope in this range is characterized by the Hurst
exponent $H$, or equivalently by the fractal dimension $D=3-H$. With this PSD,
$m_{4}$ depends heavily on the choice of the abrupt truncation at $\zeta
=\zeta_{1}$ which in the case of measured surfaces is determined by the
resolution or the sampling interval of the measuring instrument. Thus the
asperity radii continue to decrease down to atomic scales, where asperities
are defined by only a few atoms. This resolution-dependence has been much
criticized, and in the fractal limit it means that no real surface should be
sticky independently on the rms amplitude of roughness, a result which looks
paradoxical. However, we shall see that even many more recent models predict
paradoxical resolution dependence of stickiness, and the subject is still controversial.

\subsection{Contact of multiscale surfaces without adhesion}

\label{S-multiscale}

Curiously, a true multiscale model of rough surfaces was originally discussed
(without reference to adhesion) by Archard \cite{Archard} as long ago as 1957.
Ciavarella \textit{et al.} \cite{Ciavarella2000} extended these concepts to a
true fractal profile (the Weierstrass series), and found a paradoxical fractal
`limit' in which the contact is restricted to an infinite number of
infinitesimal contact areas, each sustaining an infinite contact pressure.
This anticipated Persson's `resolution-dependent' solution \cite{Persson2001}
of the adhesionless rough contact problem which at low nominal pressures
$p_{\mathrm{{nom}}}$ predicted that the total actual contact area $A_{c}$ is given
by
\begin{equation}
\frac{A_{c}}{A_{nom}}\sim\frac{p_{\mathrm{{nom}}}}{E^{\raisebox{0.7mm}{$ *$}}%
}\frac{1}{\sqrt{m_{2}}} \label{Persson2}%
\end{equation}
where $A_{nom}$ is the `nominal' or `apparent' contact area and $m_{2}$ is the
second moment of the height PSD which also coincides with the slope variance.
This dependence on rms slope is sensitive to the PSD truncation, so [e.g.]
$A_{c}\rightarrow0$ as $\zeta_{1}\rightarrow\infty$ in eq. (\ref{eqPSD})
\cite{CiaPap2017}. Surprisingly, an identical result is obtained from asperity
models \cite{Bush} except for the exact prefactor, and this result caused
quite a discussion in the literature \cite{Putignano, Carbone2008}.

An arguably more important conclusion for adhesionless contact due to Persson
\cite{Persson2007} is that some macroscopic relationships, notably that
between load and displacement, tend to a converged result in the fractal
limit, whereas asperity model theories remain ill-posed as a result of
neglecting interaction effects unless these are introduced numerically
\cite{Papetal2017}. Particularly rapid convergence is found for the important
case of low fractal dimension. Specifically, for $D\simeq2.2$ Persson
\cite{Persson2007} gives
\begin{equation}
\frac{p_{\mathrm{{nom}}}\left(  g\right)  }{E^{\raisebox{0.7mm}{$ *$}}}%
\simeq\zeta_{0}h_{\mathrm{{rms}}}\exp\left(  \frac{-\bar{g}}{\gamma
h_{\mathrm{{rms}}}}\right)  \label{Persson1}%
\end{equation}
where $\bar{g}$ is the mean separation, $\gamma\simeq0.5$ and
$h_{\mathrm{{rms}}}=\sqrt{m_{0}}$ is the rms height, which depends only weakly
on the truncation $\zeta_{1}$. Notice that the height variance $m_{0}$
coincides with the zeroth moment of the PSD. Another relationship that is only
weakly dependent on fine scale roughness is that between electrical contact
resistance and nominal pressure \cite{Bounds}.

Equations (\ref{Persson1}, \ref{Persson2}) are exemplary of two types of
result in the contact of fractal rough surfaces: those that are determined
primarily at the coarse scale and that are therefore not sensitive to
measurement resolution, and those that are not convergent and indeed that give
paradoxical predictions when ultrafine scale features are included.

\subsection{DMT-type solutions}

We first remark that both JKR and DMT solutions of Sections \ref{SJKR} and
\ref{SDMT} respectively retain a dichotomy between regions of contact and
separation and hence can be expected to give results that are sensitive to the
truncation limit $\zeta_{1}$. Of the two approaches, the DMT method seems to
be the more appropriate, since we anticipate large numbers of small contact
areas. However, it should be emphasized that the Tabor parameter of Section
\ref{STabor} cannot be directly applied to rough surface contact. Attempts to
define a generalized Tabor parameter for rough surface contact
\cite{Scaraggi,Pastewka,CiavarellaXu} generally predict that this parameter
will tend to zero in the fractal limit [$\zeta_{1}\rightarrow\infty$], seeming
to imply that a rigid-body [Bradley] solution would be appropriate in this
limit. However, elastic deformation occurs on all length scales and is not
rendered negligible by the presence of additional arbitrarily short wavelength roughness.

Persson and Scaraggi \cite{Scaraggi} developed a DMT-type solution for contact
of nominally flat rough surfaces, where the adhesionless Persson's solution
could be used since it contains an approximate expression for the probability
distribution $\Phi(g)$ for the local gap $g$ between the surfaces in regions
of separation which could then be convoluted with any desired
traction-separation law to obtain the nominal traction, as in the `force'
version of the DMT theory. Their results show a large dependence on the exact
shape of the force-separation law when amplitude of roughness is low, which is
to be expected since in the limit of no roughness, the traction-separation law
itself should be re-obtained. However, comparison with a previous (JKR-based)
theory \cite{Persson2002} seem to indicate large discrepancies, and there was
no detailed investigation of the effect of increasing truncation at
$\zeta=\zeta_{1}$ .

\subsubsection{Numerical solutions}

Persson and Scaraggi compared their theoretical predictions with a numerical
solution using a `Green's Function Molecular Dynamics' [GFMD] algorithm, a
method which has been extensively used for such studies. Essentially, the
elastic deformation is related to the discretized normal traction using an
appropriate Green's function, but the resulting set of non-linear equations at
nodal points is solved using a molecular dynamics algorithm. Solutions are
typically obtained over a rectangular grid with initial nodal heights chosen
to approximate a surface with the PSD of equation (\ref{PSD}). However,
computational considerations place limits on the practical mesh refinement, so
that even the most sophisticated codes such as those of Pastewka and Robbins
\cite{Pastewka} and that used in M\"{u}ser's recent `Contact Challenge'
\cite{Muser} can only describe surfaces with PSDs spanning about three decades
--- e.g. nanometer to micrometer scales.

Pastewka and Robbins \cite{Pastewka} developed a numerical model in which the
nodes are identified with the atoms at the plane [100] surface of an fcc
crystal in a square region with sides in the range $512a_{0}$ to $8192a_{0}$,
the latter corresponding to around one micrometer with typical interatomic
spacings $a_{0}$. Periodic boundary conditions were imposed at the edges of
the modelled region and the interaction of surface atoms with a rigid
indenting rough surface was governed by a force law approximating [and
truncating] the Lennard-Jones law \cite{LJ}.

\subsubsection{Pastewka and Robbins' theory}

In their numerical solution, Pastewka and Robbins define the contact area
$A_{\mathrm{{rep}}}$ as comprising those atoms experiencing repulsive
[compressive] forces and observe that at small applied loads $N$, it remains
approximately linear with load [compare with the adhesionless equation
(\ref{Persson1})] even in the presence of adhesion. Indeed the morphology of
the regions defining $A_{\mathrm{{rep}}}$ at appropriate force levels was
found to be only weakly dependent on adhesion, suggesting that we are in the
DMT r\'{e}gime. If this is assumed to be exactly true, we can construct a DMT
solution by (i) finding the relation between $A_{\mathrm{{rep}}}$ and $N$
without adhesion, and then (ii) modifying $N$ by summing the tensile tractions
in regions close to the perimeter of $A_{\mathrm{{rep}}}$. Based on their
numerical observations, Pastewka and Robbins estimate this correction by
assuming the existence of a `boundary layer' of tractions, leading to a term
proportional to the perimeter, and on the basis of this calculation they
define a criterion for `stickiness' such that as we move from the condition
$A_{\mathrm{{rep}}}=0$ [complete separation], the normal load initially
becomes tensile, implying that there must be another state with $N=0$ but
$A_{\mathrm{{rep}}}>0$. They also make the interesting observation that the
fractal dimension of $A_{\mathrm{{rep}}}$ is the same as that of its perimeter.

Pastewka and Robbins' stickiness criterion contains the slope and curvature
variances of the surface and if these parameters are expressed in terms of the
PSD (\ref{PSD}), it can be shown that stickiness requires that
\begin{equation}
\zeta_{1}^{(1-5H/3)}<C\;,\label{PR2}%
\end{equation}
where $C$ is a positive constant. Hence, in the fractal limit $\zeta
_{1}\rightarrow\infty$, \textit{all} surfaces with $D<2.4$ should be sticky,
and no surfaces with $D>2.4$ should be sticky. This conclusion seems quite
counterintuitive and we shall find it in contrast with two more recent
theories we discuss in the next two sections, which define stickiness based on
"pull-off". There seems to be a possibly conflict in the definition of
stickiness, but recent investigations \cite{violano} have clarified that this
is contradicted even by Persson and Scaraggi's model \cite{Scaraggi}.

\subsubsection{BAM solution}

\label{S-BAM-rough}

Ciavarella \cite{BAM} applied the BAM approximation [discussed in the context
of the spherical contact problem in Section \ref{S-BAM}] to the case of
nominally flat bodies with Gaussian random roughness. This approach has the
advantage of resulting in a closed form solution and has similarities with
DMT-like models in that some parts of the repulsive solution are used for the
adhesive problem. However, rather than estimating local separations and
convoluting them with a given force-separation law (as done by Persson and
Scaraggi \cite{Scaraggi}), BAM assumes the simplified Maugis-Dugdale
force-separation law and makes an independent estimate for the repulsive and
adhesive components of the load. The results \cite{BAM} show that the pull-off
traction is principally determined by $h_{rms}, \zeta_{0}$ and becomes
independent of the short wavelength truncation $\zeta_{1}$, as in the
adhesionless load-separation relation (\ref{Persson1}). Therefore, we can
easily construct counterexamples to the predictions of the Pastewka-Robbins
criterion (\ref{PR2}).

\subsubsection{Joe and Barber's theory}

\label{S-Joe}

Persson's adhesionless theory \cite{Persson2001} tracks the evolution of the
probability distribution $\Phi(p)$ for contact pressure as infinitesimal
increments of the roughness PSD are added. Joe and Barber \cite{Joe2017}
adapted this approach using instead the distribution $\Phi(g)$ of the local
gap $g$, the evolution of which was determined by performing a linear
perturbation on the interaction between elastic deformation and the
Lennard-Jones force law. The results agree well with those from the GFMD
algorithm due to Persson \& Scaraggi \cite{Scaraggi} for a relatively
narrow-band PSD, and in particular they exhibit convergence of the relation
between mean traction $\bar{\sigma}$ and mean gap $\bar{g}$ as the truncation
wavenumber $\zeta_{1}$ is increased without limit.

\begin{figure}[th]
\centering
\includegraphics[height=60mm]{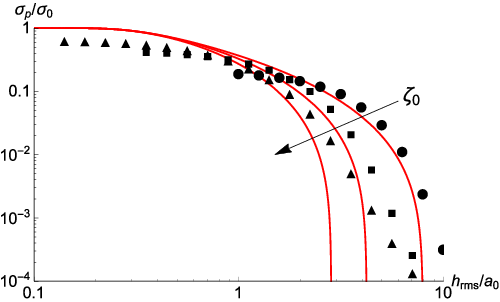}\caption{Comparison between BAM
[lines] and Joe and Barber's theory [points] for the normalized pull-off
traction as a function of height standard deviation for the three cases
$E^{*}\varepsilon\zeta_{0}/\sigma_{0}=0.053, 0.213$ and
$0.53$.}%
\label{BAM_JOE}%
\end{figure}

This method can be applied only to relatively fine-scale [short wavelength]
roughness, since longer wavelengths interacting with the Lennard-Jones force
law exhibit instabilities of the kind discussed in Section
\ref{S-instabilities}. However, if a broader PSD is partitioned into two
tranches, the theory can be used to determine the relation $\bar{\sigma}%
(\bar{g})$ for the fine-scale tranche alone, and the complete contact problem
can then be regarded as one for the coarse-scale tranche alone, but with
$\bar{\sigma}(\bar{g})$ functioning as a modified force law that accounts for
the presence of the fine scale roughness. Joe and Barber \cite{Joe2018}
exploited this idea using an iterative approach to predict the relation
$\bar{\sigma}(\bar{g})$ for broadband PSDs of the form (\ref{eqPSD}). They
also presented contour plots for pull-off traction and effective [reduced]
interface energy as functions of $m_{0}$ and the lower truncated wave number
$\zeta_{0}$ for PSDs without the `roll-off' range below $\zeta_{0}$.
Sinusoidal instabilities are still possible [and indeed are physically
reasonable] for relatively smooth surfaces with long wavelength content, but
in the stable range, the pull-off traction is predominantly determined by the
height standard deviation $h_{\mathrm{{rms}}}$ or equivalently $m_{0}$. Fig.
\ref{BAM_JOE} shows a comparison between the prediction of pull-off traction
for three different values of $\zeta_{0}$ using this theory [points] and using
the BAM approximation from Section \ref{S-BAM} [lines].

\subsection{JKR models}

If we substitute typical a asperity radius for $R$ in the definition of the
Tabor parameter (\ref{mu}), the resulting value will generally imply that a
JKR formulation is inappropriate for rough surface contact, unless an
unrealistically small value is taken for the truncation wavenumber $\zeta_{1}%
$. With this in mind, Maugis \cite{Maugis} modified Fuller and Tabor's theory
\cite{FullerTabor} to replace the JKR force-displacement law for an individual
asperity by a DMT-Maugis law, and showed that the predicted adhesion reduction
was still dependent on the parameter $\theta$ of equation (\ref{FT}), with
very minor differences. This is however most likely due to the very strong
assumption of independent asperities behaviour.

Despite the limitation to coarse-scale roughness, the mathematical simplicity
of the JKR approach has tempted many authors to use it, but reported results
need to be interpreted with caution. Persson and Tosatti \cite{PerssonTosatti}
defined an effective interface energy as the difference between the
theoretical interface energy $\Delta\gamma$ and the elastic strain energy in a
state of full contact. This is essentially the work per unit area needed to
separate the interface from a state of full contact. The elastic interface
energy for full contact is unbounded for surfaces with fractal dimension
$D\geq2.5$, so this theory would predict that such surfaces could never
adhere, even for arbitrarily small height variance $m_{0}$. By contrast, for
$D<2.5$ full contact is predicted to be possible regardless of $m_{0}$. These
results are inconsistent with the results of Joe and Barber \cite{Joe2017} and
the BAM theory \cite{BAM}. Persson and Tosatti also considered a possible
enhancement of effective interface energy due to the fact that the area of a
rough surface exceeds that of the projected plane, but if slopes are
sufficiently large to make this term significant, the original basis of the
calculation of $\Delta\gamma$ from the interaction of adjacent atoms is itself
questionable. In a later paper \cite{Persson2002}, Persson adapted his
adhesionless contact theory by modifying the boundary condition at zero
traction to include a scale-dependent finite \textit{detachment stress }[the
meaning of this stress is not very clear, and seems related to a "remote"
stress rather than a local one: clearly, in a JKR model, infinite negative
stresses should be allowed]. The predictions of these theories show both
qualitative and quantitative discrepancies relative to numerical solutions,
even when the latter are based on the JKR assumptions
\cite{Carbone2009,Carbone2015}.

Afferrante et al. \cite{Afferrante2015} extended the Weierstrass-Archard model
\cite{Ciavarella2000} to include JKR adhesion, and found a similar conclusion:
namely that for low fractal dimension, the contact area converges to a finite
limit and full contact can occur at all scales. On the other hand, Ciavarella
\cite{CiaJKR2018} showed that the [approximate] generalized JKR solution of
Section \ref{genJKR} introduces a dependence on surface slopes even in the
load-separation relationship, and predicts no adhesion in the fractal limit
for all fractal dimensions, contrary to the theories of Sections
\ref{S-BAM-rough}, \ref{S-Joe}. The JKR formulation also leads to erroneous
conclusions when applied in the limit `almost complete contact', as we shall
see in the next Section.

\subsubsection{Almost complete contact}

In Tribology it is conventional to assume that the actual contact area is much
smaller than the nominal area --- in other words that $A_{c}/A_{\mathrm{{nom}}}%
\ll1$ in equation (\ref{Persson2}). This view dates back to the pioneering
studies of friction due to Bowden and Tabor \cite{BT}, who argued that full
contact ($A_{c}=A_{\mathrm{{nom}}}$) is impossible, at least for rough metal
surfaces, due to work hardening. The condition $A_{c}/A_{\mathrm{{nom}}}\ll1$ is
also a fundamental requirement for asperity models, since these clearly only
make sense when contact is restricted to the highest points on the surface.
However, with increased interest in flexible materials such as rubber and
polymers, the possibility of larger contact ratios must be considered, even in
the elastic r\'{e}gime, including cases where actual contact occurs everywhere
except at the deepest depressions.

A good starting point for this discussion is Johnson's JKR solution
\cite{Johnson-Westergaard} for the partial contact of bodies with one- and
two-dimensional sinusoidal profiles. Johnson first determines the elastic
contact traction required for full contact and then constructs the partial
contact solution by superposing a correction comprising a set of `pressurized
cracks' opened by pressures equal and opposite to the tractions at full
contact. This idea was extended by Xu \textit{et al. } \cite{Xu2014} for a
rough surface in the absence of adhesion. In the case of full contact, the PSD
for contact traction can be written down in terms of that for surface heights
[e.g. equation (\ref{eqPSD})]. In particular, the moments $m_{n}^{p}$ of the
traction PSD are related to those of the height PSD through
\begin{equation}
m_{0}^{p}=\frac{1}{2}E^{\ast2}m_{2},\qquad m_{2}^{p}=\frac{1}{3}E^{\ast2}%
m_{4},\qquad m_{4}^{p}=\frac{3}{10}E^{\ast2}m_{6} \label{mp}%
\end{equation}
\cite{Ciavarella2015,Xu2017}, and Nayak's random surface theory \cite{Nayak}
can then be used to determine the distribution and properties of tensile
'peaks' of this distribution. Each of these peaks defines a possible
separation region and the total separation area is then determined by a
summation analogous to that used in classical asperity model theories.

We note from equation (\ref{mp}) that the moments $m^{p}_{n}$ of the traction
distribution are related to higher moments $m_{n+2}$ of the height
distribution, which suggests that the results might be very sensitive to the
truncation limit $\zeta_{1}$. However, results show that the normalized total
separation area depends only on the Nayak bandwidth parameter
\begin{equation}
\alpha^{p}=\frac{m_{0}^{p}m_{4}^{p}}{(m_{2}^{p})^{2}}=\frac{27}{20}\frac
{m_{2}m_{6}}{m_{4}^{2}}%
\end{equation}
for the traction PSD, which is only weakly dependent on the truncation. Also,
Ciavarella \cite{Ciav2016full} has shown that if a bearing-area argument is
used to estimate the separation area from the full contact pressure, and if a
corrective factor 4/3 is used for the this area, the Xu model leads exactly to
Persson's well-known solution \cite{Persson2001}. A very detailed numerical
investigation is difficult under almost full contact, but the comparisons in
Ciavarella \cite{Ciav2016full} seem to indicate that Persson's solution,
although asymptotically correct in full contact, may indeed more less accurate
than the `traction asperity' theory.

Ciavarella \cite{Ciavarella2015} extended this method to include adhesion
using the JKR approach, and found that there is now a dependence on $m_{6}$
independently of the bandwidth parameter. For low fractal dimensions, adhesion
enhancement in the form of larger and larger contact area seems to be obtained
as $\zeta_{1}$ is increased. However, if the more realistic Maugis-Dugdale
traction law is assumed in the separated region, a transition to a
non-hysteretic r\'{e}gime is found \cite{CiavarellaXu}, depending on the rms
surface slope $\sqrt{m_{2}}$. Hence, in the fractal limit, the contact
normalized contact area tends to the value without adhesion. This transition
can be characterized by a generalized Tabor parameter, where however the
process zone dimension $s_{0}$ of equation (\ref{s0}) is compared with
characteristic dimensions of distinct separation areas, rather than of contact areas.

\subsection{Tabor parameter for multiscale surfaces}

For the contact of spheres, we have seen that the exact solution [using the
Lennard-Jones traction law] is well described by the JKR theory when the Tabor
parameter $\mu$ is large, and by the DMT theory [or even Bradley's rigid body
solution] when $\mu$ is small. Several authors
\cite{Scaraggi,Pastewka,CiavarellaXu} have defined `scale-dependent' Tabor
parameters, all of which tend to zero in the fractal limit $\zeta
_{1}\rightarrow\infty$. A na\"{\i}ve comparison with the sphere problem would
then suggest that the solution in this limit could be obtained by assuming the
contacting bodies to be rigid, but this is clearly in error, since a Gaussian
surface has no highest point, so a rigid-body solution would imply infinite
separation. By contrast, theories based on an interfacial traction law, such
as those described in Sections \ref{S-BAM-rough}, \ref{S-Joe} exhibit
progressively weaker dependence on $\zeta_{1}$ and indeed converge on a
meaningful result in the fractal limit. Thus, although we can argue
rigorously, as in Section \ref{STabor}, that the JKR solution tends
asymptotically to the exact solution when $\mu\rightarrow\infty$, no such
proof exists for the case where $\mu\rightarrow0$.

In general, both DMT and JKR solutions (including numerical solutions e.g.
\cite{Carbone2009,Carbone2015}) for the contact of randomly rough surfaces
should be regarded with some caution, not to say skepticism, since the
appropriateness of these approximations depends on parameters which are often
not well characterized, such as the smallest width of a representative contact
area. At present there is no well-defined `map' of the regions of
rough-surface parameter space in which these theories might reasonably be
applied, not least because numerical solutions are computationally demanding
and hence necessarily limited in scope. In this regard, the iterative approach
used by Joe and Barber \cite{Joe2018} defines a method for spanning a broad
spectrum PSD without necessitating a choice between the two classical approximations.

\subsection{Adhesion enhancement}

\label{S-enhancement}

Experiments by Briggs \& Briscoe \cite{Briggs} with rough perspex cylinders
rolling on a flat rubber surface showed an interesting result: adhesion energy
apparently \textit{increased} with submicron roughness amplitude as compared
with the nominally smooth case. In pure rolling, the resistance must be a
combination of viscous losses and adhesive hysteresis, but these authors also
reported cases the results of pull-off experiments some of which also showed
enhancement due to roughness.

\begin{figure}[th]
\centering
\includegraphics[height=65mm]{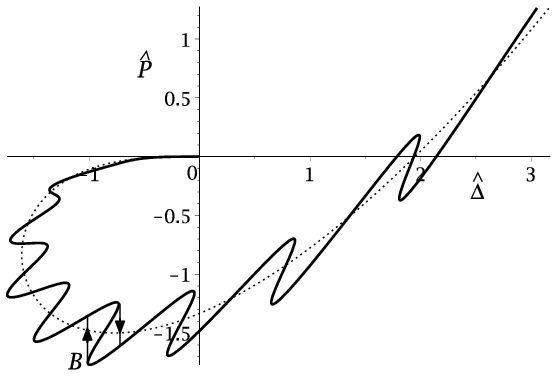} \caption{Force-indentation curve for
a sphere with axisymmetric waves (solid curve) compared with the classical JKR
curve (dotted line). Due to unstable jumps the dissipated energy is highly
increased for the wavy surface (from \cite{Guduru}).}%
\label{FigGuduru}%
\end{figure}

We recall that Persson and Tosatti \cite{PerssonTosatti} have suggested that
enhancement may result from the fact that the area of a rough surface exceeds
that of the projected plane. An alternative model was suggested by Guduru
\cite{Guduru} who considered the contact of a sphere with a rough surface
modelled as a set of concentric waves. Using the JKR model and assuming that
the contact area is simply connected (i.e. a circle), he obtained the
load-indentation plot of Fig. \ref{FigGuduru}, which exhibits oscillations
about the `smooth' curve (shown dotted) as the sphere contacts with each
successive wave. The maximum tensile force occurs at $B$ and clearly exceeds
the value for a smooth sphere, but also loading and unloading under
displacement control involve unstable jumps, such as those indicated by the
arrows in Fig. \ref{FigGuduru}, implying that toughness as well as pull-off is
enhanced. Kesari and Lew \cite{Kesari2011} provided an elegant solution for
the envelope of Guduru's curve, and Ciavarella \cite{Cia2016} showed that in
fact Kesari's solution corresponds to an increased value of interface energy
on unloading (whereas it is reduced during loading), which depend on the
parameter introduced by Johnson \cite{Johnson-Westergaard} to characterize the
adhesive contact of bodies with two-dimensional sinusoidal roughness. These
models predict that enhancement continues to increase with larger roughness
amplitudes, which is clearly unrealistic. In particular, a simply-connected
contact area would then be difficult to achieve, even with a large preload,
and the corresponding spatial oscillations would imply the existence of
extended regions where the tensile tractions exceed $\sigma_{0}$, thus
invalidating the JKR assumption.

Kesari et al. \cite{Kesari-etal} investigated a more realistic Gaussian form
of microscale roughness, and reported AFM experiments showing a marked
difference between loading and unloading, with the force on unloading
depending on the maximum indentation depth (depth-dependent hysteresis (DDH)).
Deng \& Kesari \cite{Deng} have also suggested that energy dissipation comes
from two sources, one independent of maximum indentation depth and the other
mainly due to roughness, which for large enough roughness amplitude
essentially comprises dissipation in asperity contacts.

\subsection{Effect of roughness on hysteresis}

If in a contact problem the load-displacement relation is multi-valued, then
different branches will generally be followed during loading and unloading,
resulting in hysteretic energy loss. We have reported several cases of this
kind in the preceding pages, including for example in Sections \ref{uniform},
\ref{S-pattern}, and \ref{S-enhancement}. For the sphere, hysteresis is
predicted using the JKR approximation, but not with the DMT approximation.
Indeed, the DMT solution of any geometrically defined problem predicts a
single-valued relation between force and displacement, since we first solve
the adhesionless problem, which has a unique solution, and we then modify the
resulting force by a convolution of the interface-traction law with the gap
from this same unique solution.

It is then tempting to argue that since fine-scale roughness generally pushes
us into the DMT regime, rough surface contact should not involve hysteresis.
However, this is not the case. Rough-surface contact theories generally (i)
assume that the nominal pressure is statistically uniform over an infinite
area, and (ii) eliminate the `zeroth-order' [uniform] elastic deformation of
the contacting bodies, since for half spaces, this would be infinite except in
the special case of an incompressible material. The uniform term is also
generally eliminated in numerical models. In effect, these theories generate
the properties of a fictitious `non-linear layer' which, if attached to the
surface of a smooth body, would then mimic the effect of roughness in the
actual body. This process essentially decouples the roughness scale from the
necessarily finite dimensions of the actual contacting bodies and can be seen
as a form of homogenization.

The mechanical behaviour of the layer can be described by a relation
$\bar{\sigma}(\bar{g})$ between mean traction and mean gap, which would reduce
to [e.g.] the Lennard-Jones law in the limit of vanishing roughness, but which
in general will have a lower maximum [pull-off] traction and a larger range of
effectiveness \cite{Joe2018}. As long as $\bar{\sigma}(\bar{g})$ exhibits a
tensile range, there must be a range with negative slope, so instabilities and
associated multi-valued force-displacement relations are possible under force
or displacement control. These may involve `jumps' between states with uniform
tractions, or the development of patterns as discussed in Section
\ref{S-pattern}.

In discussing Figure \ref{hysteresis}, we noted that solutions even of the
smooth contact problem are very sensitive to numerical perturbations in and
just outside the unstable range. This effect is of course even more pronounced
if the surfaces have random roughness, provided this is not of such large
amplitude as to suppress the instability. The patterns developed analogous to
those of Figure \ref{pattern} are now irregular and significant differences
are observed between different realizations of the same roughness statistics.

If the JKR approximation is used to describe the roughness scale, the relation
$\bar{\sigma}(\bar{g})$ may itself be multi-valued, as in the case of the
Fuller and Tabor model. Greenwood \cite{Greenwood2017} showed that the
unloading curve then depends on the maximum compressive traction achieved
during loading, which is a form of history-dependence. Guduru's solution [see
Section \ref{S-enhancement}] exhibits similar behaviour. However, since length
scales cannot be too small for the JKR approach to be valid, there is some
question as to whether the micro- and macro-scales can then be effectively
decoupled. This question also arises in the broad spectrum results of Joe and
Barber \cite{Joe2018} which show an unstable range at small lower wavenumber
$\zeta_{0}$ that might reasonably be interpreted as a macroscale effect.

\subsection{Effect of plasticity in rough surface contact}

In Section \ref{S-multiscale}, we showed that for the elastic contact of
multiscale rough surfaces without adhesion, the predicted actual contact area
decreases without limit as $\zeta_{1}$ increases, implying a corresponding
increase in mean contact pressure. Clearly this process must eventually be
limited by inelastic effects. Gao and Bower \cite{Gao} extended the
Weierstrass solution of Ciavarella \textit{et al. } \cite{Ciavarella2000} to
an elastic rigid-plastic material and showed that in this case the total
actual contact area tends to a finite limit, corresponding to a mean contact
pressure close to $6\sigma_{Y}$. Similar results were reported by Pei
\textit{et al. } \cite{Pei} based on a finite-element solution for a random
rough surface, and these authors also noted an approximately linear relation
between total contact area and force --- a result that was postulated by
Bowden and Tabor \cite{Bowden} as early as 1939.

The present authors are unaware of any investigation of the contact of
elastic-plastic rough bodies including adhesion, but based on the arguments in
Section \ref{S-plasticity}, we might anticipate that plastic deformation would
increase the `comformability' of the surfaces and hence increase [for example]
the pull-off traction, and make it strongly \textit{load-dependent }[that is,
in addition to the load-dependence which may already arise in elastic adhesive
contacts]. In support of this claim, we note that Mesarovic and Johnson
\cite{Mesarovic} predict an increase in pull-off force due to plasticity for a
single sphere, and their results might reasonably be incorporated in a
modified asperity model theory. Also, Pei \textit{et al. } \cite{Pei} reported
an increase in the total actual contact area during unloading relative to that
during loading which would also conduce to increased pull-off. However, notice
that roughness spectra generally extend to the nanometer scale and on this
length scale conventional plasticity may not provide a good description of the
inelastic material behaviour, and this itself needs to be investigated more.

\section{Bio-inspired adhesion}

While we have so far concentrated in theories about idealized geometries,
Nature has developed efficient mechanisms to adhere to almost any kind of
surface with a lot more freedom of choice, and indeed theoretical solutions
based on a perfectly homogeneous, nominally flat geometry fails to explain
many of these effects. Indeed, non-patterned surfaces exhibit too weak
adhesion capabilities, because roughness stress concentrations and defects
easily destroy the effect of interface energy. What makes bio-adhesive systems
exceptional is their anisotropy, and their self-cleaning and wear resistance
properties. The literature on this topic has largely expanded in recent years,
and here we shall cover only a very limited amount of material, far less
detailed than in some of the previous topics.

Synthetic pressure sensitive adhesives commonly used in domestic or industrial
applications deliver either strong or weak adhesion, but require similar
energy for detachment. Strong synthetic adhesive are difficult to detach,
whereas weak adhesives detach easily. Systems of this kind are unsuitable for
locomotion. By contrast, a gecko is able to sustain several times its weight,
but is also able to detach its foot in 15 ms and with negligible detachment
force \cite{Autumn2006, Hensel2018}. Gecko adhesion relies on non-specific van
der Walls forces \cite{Autumn2002} and its pads are covered by millions of
hairy set\ae \/ of characteristic dimension of hundred microns, which split in
finer endpoints, called spatul\ae , of nanometric dimension, leading to a
multiscale hierarchical structure.

One of the keys to this impressive performance is indeed `contact splitting'.
Hensel \textit{et al. } \cite{Hensel2018} found that in general, the pull-off
force $P_{n}$ for a micropattern with $n$ contacts is related to that for a
contact without splitting, $P_{0}$, by%
\begin{equation}
P_{n}=n^{s}P_{0}%
\end{equation}
where $s$ is called the `contact splitting efficiency', which for
hemispherical tips is equal to $1/2$. Further, if one applies fracture
mechanics arguments to the detachment of an elastic flat-ended pillar of
diameter $D_{p}$ perfectly bonded to a rigid substrate, the energy release
rate is found as
\begin{equation}
G\sim\frac{\sigma}{E}D_{p}^{0.81}l^{0.19}%
\end{equation}
where $l$ is the length of a small crack that advances at the interface.
Equating $G$ the work of adhesion $\Delta\gamma$ leads to a pull-off stress
\begin{equation}
\sigma_{P}\sim\frac{\sqrt{E\Delta\gamma}}{D_{p}^{0.406}l^{0.094}}
\label{sigmaP}%
\end{equation}
For the length scales involved, Equation (\ref{sigmaP}) gives a scaling very
close to that for a crack $(-1/2)$, and indeed the sum of the two powers is
$1/2$. The weak power in crack length $l$ is due to the assumption of
$l<<D_{p}$. When the pillar diameter is small, there is a cohesive failure
r\'{e}gime, with detachment occurring at the theoretical strength of the
material $\sigma_{0}$. Notice that this analysis predicts that higher modulus
pillars have higher pull-off stress, but this needs to be balanced by the
effect of roughness, for which lower modulus pillars adapt better.

\begin{figure}[th]
\centering
\includegraphics[height=60mm]{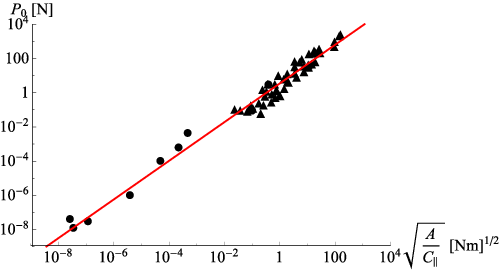} \caption{Scaling relationship for
natural and synthetic adhesives. Data align over 14 orders of magnitude. (form
\cite{Bartlett})}%
\label{scalingAoverC}%
\end{figure}

Although contact splitting provides an explanation for adhesion enhancement in
some bio applications, we are still far for a complete understanding of how
adhesion is effectively controlled at the interface. While Artz \textit{et al.
} \cite{Artz} showed a strong correlation in flies, beetles, spiders, and
lizards between the areal density of attachment hairs and the body mass, later
Peattie and Full \cite{Peattie} surveyed $81$ species with hierarchical
fibrillar structures and found no such correlation when the data are analyzed
within the same taxa (see also \cite{Webster}). Indeed Bartlett \textit{et al}
\cite{Bartlett} proposed a more general criterion based on total energy
minimization, which predicted that the maximum adhesion force would scale
according to the relationship%
\begin{equation}
P_{0}\sim\sqrt{\Delta\gamma}\sqrt{\frac{A_{c}}{C_{\parallel}}} \label{Bart}%
\end{equation}
where $A_{c}$ is the area of intimate contact and $C_{\parallel}$ is the compliance in the
direction of the applied load. Figure \ref{scalingAoverC} shows an impressive
correlation between equation (\ref{Bart}) and results for both natural and
synthetic adhesives over 14 orders of magnitudes.

Since $\Delta\gamma$ is an uncontrollable parameter that depends on the
interface properties, an optimal design requires that we maximize the ratio
$A_{c}/C_{\parallel}$, but this is not an easy task. The system must be soft enough to
maximize the contact area and adapt well on any kind of substrate, but also
rigid enough in the direction of the applied load. In this respect,
hierarchical fibrillar structures are likely to be one of the more promising
designs (perhaps the most successful), but not the only one. Indeed, Bullock
\textit{et al. } \cite{Bullock} compared hairy (Gastrophysa viridula) and
smooth (Carausius morosus) pads finding comparable adhesive performance,
except that hairy pads can exhibit more anisotropy due to the ability to
control each seta individually.

\subsection{Patterned surfaces}

Inspired by biological design solutions, micro- and nano-patterned surfaces
have been developed. These rely on pillars or dimples of various shapes, whose
individual small scale makes it possible to have them defect-tolerant and
reach high values of strength, close to $\sigma_{0}$. Several authors have
attempted to replicate solutions in Nature with micropatterned dry adhesives
to obtain reversible capabilities to grip, position, and release objects
\cite{Akerboom2015, Kamperman, Hwang, King, Benz}.

For an arrangement of pillars, a simple argument is that the elastic strain
energy stored in a single pillar is effectively dissipated during pull-off
\cite{Hensel2018}. The remaining load then has to be redistributed over the
remaining surface and the crack needs to nucleate again at the next pillar for
pull-off to proceed. Hence `crack trapping' enhances the effective work of
separation at the next pillar by $\pi D^{2}\sigma_{P}^{2}L/2E$, where $L$ is
the pillar height. Therefore, a small elastic modulus with long pillars having
high individual pull-off stresses is beneficial for this effect to be maximized.

Barreau and co-authors \cite{Barreau} have shown that one strategy for better
adaptation to surface roughness is to use a small pillar diameter to take
advantage of the contact splitting effect, but not smaller than the mean
spacing between local peaks on the substrate, insofar as this can be defined
for a multiscale surface. The problem is that when diameter is too small
compared with this criterion, bending and buckling events occur, storing
strain energy, which effectively reduce adhesion.

An alternative strategy is to design the contact geometry to include mushroom
or funnel-shaped tips. These shapes can increase the pull-off stress even by
an order of magnitude because of the shift from the severe singular edge
stresses of the flat punch to the more uniform stresses with almost no
singularity at the edge of the mushroom flaps \cite{Balijepalli}. More
precisely, the singular stress multiplier is determined by the thickness of
the flaps, rather than the diameter of the fibril.

\begin{figure}[th]
\centering
\includegraphics[height=65mm]{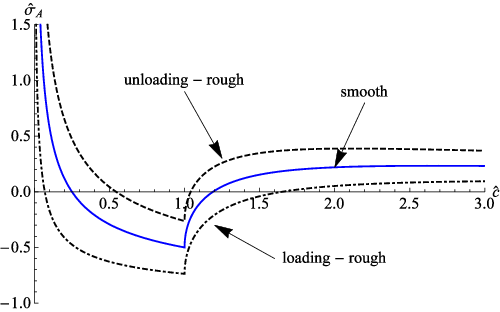} \caption{Loading curves for smooth
(solid lines) and rough dimple (dashed lines).}%
\label{DimpleJKR}%
\end{figure}

For surfaces with nano/micro-dimples, McMeeking \textit{et al. }
\cite{McMeeking} proposed an elegant model comprising essentially the JKR
solution for a single depression in one of the surfaces. Interestingly a
bi-stable pressure-sensitive adhesive mechanism was obtained, with two
distinct states of weak or strong adhesion. As in Johnson's solution for the
sinusoid \cite{Johnson-Westergaard}, the model suffers from the weakness that
with the JKR assumption a theoretically infinite traction is needed to detach
the surface from a state of full contact (see the solid line in Fig.
\ref{DimpleJKR}) unless an initial perturbation such as air entrapment as
assumed. Papangelo \& Ciavarella \cite{PapCia2017} extended the McMeeking
analysis using a Maugis cohesive model to account for the adhesive
interaction. The analysis showed that the adhesive behaviour depends not only
on the work of adhesion but also on a generalized Tabor parameter $\mu$. Low
$\mu$ leads to the rigid solution which shows no hysteretic behaviour, while
for large $\mu$ the cohesive solution tends to the JKR limit.

More recently, Papangelo \& Ciavarella \cite{PapCia2018} studied the effect of
an axisymmetric single-scale sinusoidal roughness superimposed on an otherwise
smooth dimple, in the JKR r\'{e}gime. As in \cite{Guduru} (see Section
\ref{S-enhancement}), the contact area is assumed to be connected, in this
case comprising the region outside a single circle. The results show that the
nominal compressive traction required to reach the full contact state is
increased by the added roughness, but when this is reached, the resistance to
pull-off is increased relative to the `smooth' case (see the dashed lines in
Fig. \ref{DimpleJKR}). In effect, the rough dimple behaves similarly to the
smooth dimple, except that the effective work of adhesion is increased during
loading and decreased during unloading. These results are analogous to those
of Guduru \cite{Guduru} for the case of a sphere with added a sinusoidal
axisymmetric waviness.

\section{Adhesion and friction}

Recently, there has been a large interest in the interplay between adhesion
and friction. In many insects, for example, it has been found that the normal
force needed to detach adhesive pads is approximately a linear function of the
shear force simultaneously applied \cite{Autumn2006, Labonte, Gravish}. This
recalls Amonton's law for friction, and indeed the model of \textquotedblleft
frictional adhesion\textquotedblright\ introduced by Autumn \cite{Autumn2006},
is a modification of the classical Amonton law, with two \textquotedblleft
friction coefficients\textquotedblright, one for compressive, and the other
for tensile loads.

Classical contact mechanics models for adhesion and friction interaction date
back to the seminal work of Savkoor \& Briggs \cite{Savkoor} who extended the
JKR solution for a smooth sphere to friction. They assumed a singular stress
field also in tangential direction (mode II), and combined the energy release
rate $G$ as
\begin{equation}
G=\frac{1}{2E^{\ast}}\left[  K_{I}^{2}+K_{II}^{2}\right]  =\Delta\gamma
\end{equation}
where $K_{I}$ and $K_{II}$ are respectively the mode I and II stress intensity factors.

However, Savkoor \& Briggs \cite{Savkoor} found that the contact area
reduction was \textit{greatly overestimated} by this "purely brittle" model,
which, in other words, assumes no frictional resistance when the crack
advances. Instead, the interface "toughness" $G_{c}$ should be considered a
function of the phase angle
\begin{equation}
\psi=\arctan\left(  \frac{K_{II}}{K_{I}}\right)
\end{equation}
and, although physical models have been advanced \cite{Hutchinson1990},
essentially a mode-mixity function $f\left(  \psi\right)  $ that includes a
fitting parameter is introduced \cite{HutchinsonSuo}
\begin{equation}
G_{c}=\Delta\gamma f\left(  \psi\right)  \label{Gpsi}%
\end{equation}

\begin{figure}[th]
\centering
\includegraphics[height=70mm]{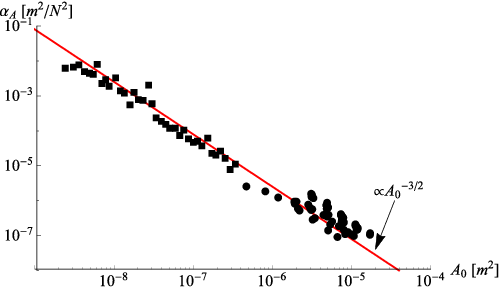} \caption{Contact area reduction
coefficient $\alpha_{A}$ as a function of the initial apparent contact area
$A_{0}$ for both sphere and rough contact. (from \cite{Sahli})}%
\label{FigPNAS}%
\end{figure}

Johnson \cite{Johnson1996} was the first to reconsider Savkoor \& Briggs
\cite{Savkoor} model adding a single empirical constant $0<\alpha<1$ to tune
the "interaction" between modes, where $\alpha=0$ corresponds to the `ideally
brittle' behavior of Savkoor \& Briggs \cite{Savkoor} which implies no
frictional resistance in the relative tangential motion of the two surfaces,
and $\alpha= 1$ corresponds to the mode uncoupling with no sensitivity to the
tangential load. Later, Johnson \cite{Johnson1997} introduced more complex
cohesive models in both mode I and mode II (mode III was removed by "averaging
around the periphery of the contact to maintain axisymmetry in the model), and
the number of constants increased. Waters \& Guduru \cite{Waters} came back on
this argument proposing their fracture mechanics model and comparing with
extensive experimental results, which showed good agreement at least until the
contact area remained circular. The possibility of cycles of slip instability
and reattachment especially for compressive normal loads appear in some
authors, which have some similarities with Schallamach waves
\cite{Schallamach}, and at the moment there is no complete understanding about
when, depending on the particular experimental testing apparatus, method and
materials used, they should appear or not (see discussion \cite{Waters}).
Finally, Sahli {et al.} \cite{Sahli} suggest a quadratic decay of the contact
area with tangential load, $A_{c}=A_{0}-\alpha_{A}T^{2}$, where $A_{0}$ is the
contact area at null tangential load $T=0$ and $\alpha_{A}$ is a fitting area
reduction coefficient. They find that the latter parameter scales with power
$-3/2$ with $A_{0}$ over about 4 orders of magnitude (Fig. \ref{FigPNAS}), for
both smooth (sphere vs plane) and rough contact and that sliding corresponds
to the condition when the contact area has reduced such that the tangential
load corresponds to the product of a material constant shear strength and the
reduced contact area.

\section{Conclusions}

We have seen that a wide range of interesting physical phenomena can result
from the interaction of adhesive [i.e. tensile] tractions between contacting
bodies and the underlying contact mechanics. These include, for example,
various kinds of instability involving jumping into and out of contact, and
also the development of self-generated patterns in systems where one would
otherwise expect a nominally uniform state. The `JKR' energetic approach
provides a convenient and reliable approximation to the solution in such cases
provided that the problem is `macroscopic' --- i.e. that the resulting contact
and separation regions are large relative to the length scale defined by
equation (\ref{s0}) --- but many modern applications of adhesive contact
mechanics fail this test, either because the contacting bodies are themselves
small, as in many biological and nano-structural problems, or because the more
or less inevitable existence of surface roughness causes the contact area to
bifurcate into a morphology defined at the nanoscale.

Various approximations have been suggested and used in these cases, including
the approximation of the traction law by a piecewise constant function, and
the so-called `DMT' approach, where it is assumed \textit{a priori} that the
region where the contact tractions are compressive is unaffected by
surrounding regions of tensile traction, which therefore only affect the
calculation of the total force. The reliability of such approaches can only be
assessed by comparison with non-linear numerical solutions using the exact
form of the interface traction law, but unfortunately computational
limitations generally restrict the dimensions of the system that can be so
modelled, typically to not exceed the micrometre scale. This problem is
particularly challenging when surface roughness is taken into account, since
roughness often spans a spectrum covering four or five orders of magnitude,
and additional complication is introduced by the fact that the surface is then
defined only in a statistical sense. We describe and compare various attempts
at treating this challenging multiscale problem.

In summary, adhesive contact mechanics is an extraordinarily rich source of
challenging problems, in which significant advances are regularly being made.

\enlargethispage{20pt}

\section*{Author contribution statement}

MC generally designed the review, JJ provided some numerical analysis for the
patterns in very smooth soft surfaces, and all authors contributed to the organization and text of the final MS.

\section*{Acknowledgements}

A.P. is thankful to the German Research Foundation (DFG) for funding the
projects HO 3852/11-1 and PA 3303/1-1.

%%%%%%%%%% Insert bibliography here %%%%%%%%%%%%%%

\section*{Nomenclature}

\begin{itemize}

\item $A_{\mathrm{{att}}},$ attractive contact area

\item $A_{c},$ total contact area 

\item $A_{nom},$ "nominal" or "apparent" contact area

\item $A_{\mathrm{{rep}}},$ repulsive contact area

\item $A_{0},$ contact area at null tangential load

\item $B(\Delta),$ bearing area

\item $C\left(  \zeta\right),$ surface heights Power Spectral Density (PSD).

\item $C_{\parallel}$, compliance in the direction of the applied load

\item $C_{0},$ power spectrum value within the roll-off bandwidth

\item $D,$ fractal dimension

\item $D_{p},$ pillar diameter

\item $E_{i},$ Young modulus

\item $E^{\ast},$ plain strain elastic modulus.

\item $G,$ energy release rate

\item $G_{c},$ interfacial toughness

\item $H,$ Hurst parameter

\item $K_{i},$ stress intensity factor

\item $L,$ pillar height

\item $P,$ compressive indentation force

\item $\hat{P},$ dimensionless compressive indentation force

\item $P_{C}(\Delta),$ compressive force in the corresponding elastic contact
problem without adhesion

\item $P_{n},$ pull-off force of a micropattern with $n$ contacts

\item $P_{0},$ pull-off force

\item $R,$ sphere radius

\item $R^{p},$ full contact pressure "asperity" radius

\item $S,$ amplitude of the sinusoidal traction distribution

\item $T$, tangential load

\item $U,$ elastic strain energy

\item $a,$ the smallest length scale associated to the geometry

\item $d_{rep},$ characteristic diameter of repulsive contact areas

\item $f\left(  \psi\right)  ,$ mode mixity function

\item $g,$ separation

\item $g_{\max},$ critical separation in Maugis potential, $g_{\max}%
=\Delta\gamma/\sigma_{0}$

\item $h,$ layer thickness

\item $h_{rms}, h_{rms}^{\prime}, h_{rms}^{\prime\prime}$ respectively surface height, slope, curvature root mean square

\item $k,$ Winkler foundation modulus

\item $k(\zeta),$ wavenumber-dependent stiffness. 

\item $k_{rep,}$ repulsive Area-load slope

\item $l,$ interfacial crack advancing at the interface

\item $l_{a}=\Delta\gamma/E^{\ast}$, characteristic adhesion length

\item $m_{n},$ n-th order moment of height PSD

\item $m_{n}^{p},$ n-th order moment of traction PSD

\item $p_{nom},$ nominal pressure

\item $s,$ contact splitting efficiency

\item $s_{0},$ length scale associated with the singular traction field

\item $u,$ elastic deformation

\item $u_{n},$ Fourier series coefficients

\item $\Gamma,$ interface energy

\item $\Delta,$ indentation depth, or (for the layer) gap that would exist
between a layer and a plane surface in the absence of elastic deformation 

\item $\hat{\Delta},$ dimensionless indentation depth

\item $\Delta\gamma,$ interface energy per unit area

\item $\Delta r,$ range of attractive forces in PR theory. \emph{ }

\item $\Pi,$ total potential energy

\item $\Omega,$ potential energy of external forces

\item $\alpha,$ interaction parameter

\item $\alpha_{A}$, fitting area reduction coefficient

\item $\beta=\frac{E\varepsilon^{2}}{h\Delta\gamma},$ dimensionless parameter
for layer instability

\item $\gamma,$ fitting parameter in Persson load-separation theory

\item $\varepsilon,$ equilibrium position

\item $\zeta=\frac{2\pi}{\lambda},$ wavenumber, with $\lambda$ the corresponding wavelength

\item $\zeta_{L},$ low wavenumber truncation

\item $\zeta_{0},$ roll-off wavenumber

\item $\zeta_{1},$ high wavenumber truncation

\item $\theta_{FT},$ Fuller and Tabor parameter

\item $\mu,$ Tabor parameter

\item $\nu_{i},$ Poisson's ratio

\item $\xi=x/h,$ dimensionless coordinate

\item $\sigma,$ traction \ [tensile positive]

\item $\sigma_{P},$ pull-off stress

\item $\sigma_{y},$ yield strength

\item $\sigma_{0},$ maximum tensile traction [theoretical strength]

\item $\psi,$ phase angle

\end{itemize}

\begin{thebibliography}{999}                                                                                              %
%\bibliographystyle{plain}
%\bibliography{<bib>}


\bibitem {LJ}J. E. Jones, (1924), On the determination of molecular fields. II.
From the equation of state of a gas, \textit{Proceedings of the Royal Society
of London}, Vol. 106 (738), pp. 463--477 [213]. doi: 10.1098/rspa.1924.0082

\bibitem {Maugis}D. Maugis, \textit{Contact, Adhesion and Rupture of Elastic
Solids}, Springer, New York (2000).

\bibitem {Bradley}R.S. Bradley, (1932). The cohesive force between solid surfaces and
the surface energy of solids. Phil Mag; 13:853--862.

\bibitem {Rumpf}H. Rumpf, \textit{Particle Technology}, Chapman \& Hall,
London (1990).

\bibitem {Rabinovich}Y.I. Rabinovich, J.J. Adler, A. Ata, R.K. Singh and B.M.
Moudgil, (2000), Adhesion between nanoscale rough surfaces: I. role of asperity
geometry; II. measurement and comparison with theory, Journal of Colloid and
Interface Science, Vol. 232, pp. 10--16, 17--24.

\bibitem {JKR} K.L. Johnson, K. Kendall, A.D. Roberts, (1971), Surface energy and the
contact of elastic solids. Proc R Soc Lond;A324:301--313. doi: 10.1098/rspa.1971.0141

\bibitem {Johnson-Westergaard}K. L. Johnson, (1995), The adhesion of two
elastic bodies with slightly wavy surfaces, \textit{International Journal of
Solids and Structures}, Vol. 32 (3--4), pp. 423--430. doi: 10.1016/0020-7683(94)00111-9

\bibitem {Popov-flat-adhesion}V. L. Popov, R. Pohrt and Q. Li, (2017), Strength
of adhesive contacts: Influence of contact geometry and material gradients.
\textit{Friction}, Vol. 5(3), pp. 308--325. doi: 10.1007/s40544-017-0177-3

\bibitem {CiaJKR2018}M. Ciavarella, (2018), An approximate JKR solution for a
general contact, including rough contacts. Journal of the Mechanics and
Physics of Solids, 114, 209-218.

\bibitem {Muller}V. M. Muller, V. S. Yushchenko and B. V. Derjaguin, (1980), On
the influence of molecular forces on the deformation of an elastic sphere and
its sticking to a rigid plane, \textit{Journal of Colloid and Interface
Science}, Vol. 77 (1), pp. 91--101. doi: 10.1016/0021-9797(80)90419-1

\bibitem {Tabor}D. Tabor, (1977), Surface forces and surface interactions, J. Colloid
Interface Sci; 58:2--13. doi: 10.1016/0021-9797(77)90366-6

\bibitem {hysteresis}M. Ciavarella, J.A. Greenwood and J.R. Barber, (2017), Effect of
Tabor parameter on hysteresis losses during adhesive contact, Journal of the
Mechanics and Physics of Solids, Vol. 98, pp. 236--244.

\bibitem {Derjaguin}B. V. Derjaguin, V. M. Muller and Yu. P. Toporov, (1975),
Effect of contact deformations on the adhesion of particles, \textit{Journal
of Colloid and Interface Science}, Vol. 53 (2), pp. 314--326. doi: 10.1016/0021-9797(75)90018-1.

\bibitem {Pashley}M. D. Pashley, (1984), Further consideration of the DMT model
for elastic contact. Colloids and Surfaces, Vol. 12 [1-2], pp. 69--77. doi: 10.1016/0166-6622(84)80090-6

\bibitem {Greenwood-DMT}J.A. Greenwood, (2007), On the DMT theory. Tribol Lett; 26:203--211. doi: 10.1007/s11249-006-9184-7

\bibitem {Maugis-1992}D. Maugis, (1992), Adhesion of spheres: The JKR-DMT
transition using a Dugdale model, \textit{Journal of Colloid and Interface
Science}, Vol. 150 (1), pp. 243--269. doi: 10.1016/0021-9797(92)90285-T

\bibitem {double-Hertz}J. A. Greenwood and K. L. Johnson, (1998), An
alternative to the Maugis model of adhesion between elastic spheres,
\textit{Journal of Physics, D: Applied Physics}, Vol. 31 (22), pp. 3279--3290.
doi: 10.1088/0022-3727/31/22/017

\bibitem {BAM}M. Ciavarella, (2018), A very simple estimate of adhesion of hard
solids with rough surfaces based on a bearing area model, \textit{Meccanica},
Vol. 53, pp. 241--250. doi: 10.1007/s11012-017-0701-6

\bibitem {Johnson1976}K. L. Johnson, (1976), Adhesion at the contact of
solids. In: Koiter, W.T. (Ed.), Theoretical and Applied Mechanics, Proc. 4th
IUTAM Congress. North Holland, Amsterdam, p. 133.

\bibitem {Mesarovic}S. D. Mesarovic, K. L. Johnson, (2000), Adhesive
contact of elastic--plastic spheres. Journal of the Mechanics and Physics of
Solids, 48(10), 2009-2033.

\bibitem {Johnson}K. L. Johnson, (1985), \textit{Contact Mechanics}, Cambridge
University Press, Cambridge.

\bibitem {Yang-comp}F. Yang, (2006), Asymptotic solution to axisymmetric
indentation of a compressible elastic thin film, \textit{Thin Solid Films}
Vol. 515 (4), pp. 2274--2283.\newline doi: 10.1016/j.tsf.2006.07.151

\bibitem {Argatov}I. I. Argatov, G. S. Mishuris and V. L. Popov, (2016),
Asymptotic modelling of the JKR adhsion contact for thin elastic layer,
\textit{Quarterly Journal of Mechanics and Applied Mathematics}, Vol. 69 (2),
pp. 161--179. doi: 10.1093/qjmam/hbw002

\bibitem {Yang-incomp}F. Yang, (2002), Adhesive contact between a rigid
axisymmetric indenter and an incompressible elastic thin film, \textit{Journal
of Physics D, Applied Physics}, Vol. 35 (20), pp. 2614--2620. doi: 10.1088/0022-3727/35/20/322

\bibitem {Papangelo2018}A. Papangelo, (2018), Adhesion between a power-law
indenter and a thin layer coated on a rigid substrate. Facta Universitatis,
Series: Mechanical Engineering, 16(1), 19-28.

\bibitem {Hannah}M. Hannah, (1951), Contact stress and deformation in a thin
elastic layer. \textit{Quarterly Journal of Mechanics and Applied
Mathematics}, Vol. 4 (1), pp. 94--105. doi: 10.1093/qjmam/4.1.94

\bibitem {shenoy2001pattern}V.~Shenoy, A.~Sharma, (2001), Pattern formation in a thin
solid film with interactions, Physical Review Letters 86~(1) 119.

\bibitem {sarkar2004patterns}J.~Sarkar, V.~Shenoy, A.~Sharma, (2004), Patterns,
forces, and metastable pathways in debonding of elastic films, Physical Review
Letters 93~(1) 018302.

\bibitem {gonuguntla2006control}M.~Gonuguntla, A.~Sharma, R.~Mukherjee, S.~A.
Subramanian, (2006), Control of self-organized contact instability and patterning in
soft elastic films, Langmuir 22~(16) 7066--7071.

\bibitem {monch2001elastic}W.~M{\"o}nch, S.~Herminghaus, (2001), Elastic instability
of rubber films between solid bodies, EPL (Europhysics Letters) 53~(4) 525.

\bibitem {Scaraggi}B. N. J. Persson and M. Scaraggi (2014), Theory of
adhesion: Role of surface roughness, \textit{Journal of Chemical Physics},
Vol. 141 (12), Art. 124701. doi: 10.1063/1.4895789

\bibitem {bi-layer}R. Mukherjee, R. Pangule, A, Sharma, and G. Tomar, (2007), Contact
instability of elastic bilayers: Miniaturization of instability patterns
\textit{Advanced Functional Materials}, Vol. 17(14), pp. 2356--2364.
doi: 10.1002/adfm.200600896

\bibitem {Vakis}A.I. Vakis, \textit{et al.}, (2018), Modeling and simulation in
tribology across scales: An overview. Tribology International, Vol. 125, pp 169-199.DOI: 10.1016/j.triboint.2018.02.005

\bibitem {FullerTabor}K.N.G. Fuller, D. Tabor, (1975), The effect of
surface roughness on the adhesion of elastic solids. Proc. R. Soc. Lond. A,
345(1642), 327-342.

\bibitem {Greenwood1966}J.A. Greenwood, J.B.P. Williamson, (1966), Proc. R.
Soc. London A 295, 300.

\bibitem {Greenwood2017}J. A. Greenwood, (2017), Reflections on and Extensions
of the Fuller and Tabor Theory of Rough Surface Adhesion. Tribology Letters,
65(4), 159.

\bibitem {McCool}J.I. McCool, (1987), Relating Profile Instrument Measurements
to the Functional Performace of Rough Surfaces, ASME J. Tribol. 109, 264-270.

\bibitem {Archard}J. F. Archard, (1957), \textquotedblleft Elastic Deformation
and the Laws of Friction,\textquotedblright\ Proc. R. Soc. London A,
243(1233), pp. 190--205.

\bibitem {Ciavarella2000}M. Ciavarella, G. Demelio, J.R. Barber, and Y.H. Jang, (2000), \textquotedblleft Linear Elastic Contact of the Weierstrass
Profile,\textquotedblright\ Proc. R. Soc. London A, 456(1994), pp. 387--405.

\bibitem {Persson2001}B.N. Persson, (2001), Theory of rubber friction and
contact mechanics. The Journal of Chemical Physics, 115(8), 3840-3861.

\bibitem {CiaPap2017}M. Ciavarella \& A. Papangelo, (2017), Discussion of
\textquotedblleft Measuring and Understanding Contact Area at the Nanoscale: A
Review\textquotedblright(Jacobs, TDB, and Ashlie Martini, A., 2017, ASME Appl.
Mech. Rev., 69 (6), p. 060802). Applied Mechanics Reviews, 69(6), 065502.

\bibitem {Bush}A.W. Bush, R. D. Gibson \& T.R. Thomas, (1975), The elastic
contact of a rough surface. Wear, 35(1), 87--111.

\bibitem {Putignano}C. Putignano, L. Afferrante, G. Carbone, \& G. Demelio,
(2012), A new efficient numerical method for contact mechanics of rough
surfaces. International Journal of Solids and Structures, 49(2), 338-343.

\bibitem {Carbone2008}G. Carbone, \& F. Bottiglione, (2008), Asperity contact
theories: Do they predict linearity between contact area and load?. Journal of
the Mechanics and Physics of Solids, 56(8), 2555-2572.

\bibitem {Persson2007}B. N. J. Persson, (2007), Relation between interfacial
separation and load: a general theory of contact mechanics. Physical review
letters, 99(12), 125502.

\bibitem {Papetal2017}A. Papangelo, N. Hoffmann, M. \& Ciavarella, (2017),
Load-separation curves for the contact of self-affine rough surfaces.
Scientific reports, 7(1), 6900.

\bibitem {Bounds}J.R. Barber, (2003), Bounds on the electrical resistance
between contacting elastic rough bodies. Proceedings of the Royal Society of
London, 459(2029), 53--66.

\bibitem {Pastewka}L. Pastewka, M.O. Robbins, (2014), Contact between
rough surfaces and a criterion for macroscopic adhesion. Proceedings of the
National Academy of Sciences, 201320846.

\bibitem {CiavarellaXu}M. Ciavarella, Y. Xu, \& R.L. Jackson, (2018), Some
Closed-Form Results for Adhesive Rough Contacts Near Complete Contact on
Loading and Unloading in the Johnson, Kendall, and Roberts Regime. Journal of
Tribology, 140(1), 011402.

\bibitem {Persson2002}B. N. J. Persson, (2002), Adhesion between elastic
bodies with randomly rough surfaces. Physical review letters, 89(24), 245502.

\bibitem {Muser}M. H. M\"{u}ser, W. B. Dapp, R. Bugnicourt, P. Sainsot,
N. Lesaffre, T. A. Lubrecht,... \& S. Rohde, (2017), Meeting the
contact-mechanics challenge. Tribology Letters, 65(4), 118.

\bibitem {Joe2017}J. Joe, M. Scaraggi, J.R. \& Barber, (2017), Effect of
fine-scale roughness on the tractions between contacting bodies. Tribology
International, 111, 52-56.

\bibitem {Joe2018}J. Joe, M.D. Thouless, \& J.R. Barber, (2018), Effect of
roughness on the adhesive tractions between contacting bodies. Journal of the
Mechanics and Physics of Solids, 118, pp. 365-373.

\bibitem {PerssonTosatti}B. N. J. Persson, E. Tosatti, (2001), The effect
of surface roughness on the adhesion of elastic solids. The Journal of
Chemical Physics, 115(12), 5597-5610.

\bibitem {Carbone2009}G. Carbone, M. Scaraggi, \& U. Tartaglino, (2009),
Adhesive contact of rough surfaces: comparison between numerical calculations
and analytical theories. The European Physical Journal E, 30(1), 65.

\bibitem {Carbone2015}G. Carbone, E. Pierro, \& G. Recchia, (2015),
Loading-unloading hysteresis loop of randomly rough adhesive contacts.
Physical Review E, 92(6), 062404.

\bibitem {Afferrante2015}L. Afferrante, M. Ciavarella, \& G. Demelio,
(2015), Adhesive contact of the Weierstrass profile. Proc. R. Soc. A,
471(2182), 20150248.

\bibitem {BT}F.P Bowden, \& D. Tabor, (1950), \textit{The Friction and
Lubrication of Solids}, Clarendon Press, Oxford.

\bibitem {Xu2014}Y. Xu, R.L. Jackson, and D.B. Marghitu, (2014), Statistical
model of nearly complete elastic rough surface contact, Int J Solids Struct,
51, 1075-1088.

\bibitem {Ciavarella2015}M. Ciavarella, (2015), Adhesive rough contacts near
complete contact. International Journal of Mechanical Sciences, 104, 104-111.

\bibitem {Xu2017}Y. Xu, \& R.L. Jackson, (2017), Statistical models of
nearly complete elastic rough surface contact-comparison with numerical
solutions. Tribology International, 105, 274-291.

\bibitem {Nayak}P. R. Nayak, (1971), Random process model of rough surfaces,
ASME Journal of Lubrication Technology, 93 (3), 98--407. doi: 10.1115/1.3451608

\bibitem {Ciav2016full}M. Ciavarella, (2016), Rough contacts near full contact
with a very simple asperity model. Trib Int, 93, 464-469.

\bibitem {Briggs}G. A. D. Briggs, \& B. J. Briscoe, (1977), The effect of
surface topography on the adhesion of elastic solids. Journal of Physics D:
Applied Physics, 10(18), 2453.

\bibitem {Guduru}P. R. Guduru, (2007), Detachment of a rigid solid from an
elastic wavy surface: theory. Journal of the Mechanics and Physics of Solids,
55(3), 445-472.

\bibitem {Kesari2011}H. Kesari, \& A. J. Lew, (2011), Effective macroscopic adhesive
contact behavior induced by small surface roughness. J. Mech.Phys. Solids 59,
2488--2510.

\bibitem {Cia2016}M. Ciavarella, (2016), On roughness-induced adhesion
enhancement. The Journal of Strain Analysis for Engineering Design, 51(7), 473-481.

\bibitem {Kesari-etal}H. Kesari, J.C. Doll, B.L. Pruitt, W. Cai, \& A.J. Lew, (2010), Role of surface roughness in hysteresis during adhesive elastic
contact. Philos. Mag. Lett. 90, 891-902.

\bibitem {Deng}W. Deng, H. \& Kesari, (2018), Depth-dependent hysteresis in
adhesive elastic contacts at large surface roughness. arXiv preprint arXiv:1803.08581.

\bibitem {Gao}Y. F. Gao, \& A. F. Bower, (2006), Elastic--plastic
contact of a rough surface with Weierstrass profile. In Proceedings of the
Royal Society of London A: Mathematical, Physical and Engineering Sciences
(Vol. 462, No. 2065, pp. 319-348). The Royal Society.

\bibitem {Pei}L. Pei, S. Hyun, J.F. Molinari, \& M.O. Robbins, (2005),
Finite element modeling of elasto-plastic contact between rough surfaces.
Journal of the Mechanics and Physics of Solids, 53(11), 2385-2409.

\bibitem {Bowden}F. P. Bowden, \& D. Tabor, (1939), The area of contact
between stationary and moving surfaces. Proc. R. Soc. Lond. A, 169(938), 391-413.

\bibitem {Autumn2006}K. Autumn, A. Dittmore, D. Santos, M. Spenko, \&
M. Cutkosky, (2006), Frictional adhesion: a new angle on gecko attachment.
Journal of Experimental Biology, 209(18), 3569-3579.

\bibitem {Hensel2018}R. Hensel, K. Moh, \& E. Arzt, (2018), Engineering
Micropatterned Dry Adhesives: From Contact Theory to Handling Applications.
Advanced Functional Materials, 1800865.

\bibitem {Autumn2002}K. Autumn, M. Sitti, Y.A. Liang, A.M. Peattie, W.R.
Hansen, S. Sponberg, ... \& R.J. Full, (2002), Evidence for van der
Waals adhesion in gecko setae. Proceedings of the National Academy of
Sciences, 99(19), 12252-12256.

\bibitem {Artz}E. Arzt, S. Gorb, \& R. Spolenak, (2003), From micro to nano
contacts in biological attachment devices. Proceedings of the National Academy
of Sciences, 100(19), 10603-10606.

\bibitem {Peattie}A. M. Peattie, \& R.J. Full, (2007), Phylogenetic analysis
of the scaling of wet and dry biological fibrillar adhesives. Proceedings of
the National Academy of Sciences, 104(47), 18595-18600.

\bibitem {Webster}N.B. Webster, M.K. Johnson, and A.P. Russell, (2009),
Ontogenetic scaling of scansorial surface area and setal dimensions of
Chondrodactylus bibronii (Gekkota: Gekkonidae): testing predictions derived
from cross-species comparisons of gekkotans. Acta Zoologica (Stockholm) 90: 18-29.

\bibitem {Bartlett} M. D. Bartlett, A. B. Croll, D. R. King, B. M. Paret, D. J.
Irschick, \& A. J. Crosby, (2012), Looking beyond fibrillar features to
scale gecko-like adhesion. Advanced Materials, 24(8), 1078-1083.

\bibitem {Bullock}J. M. Bullock, P. Drechsler, \& W. Federle, (2008),
Comparison of smooth and hairy attachment pads in insects: friction, adhesion
and mechanisms for direction-dependence. Journal of Experimental Biology,
211(20), 3333-3343.

\bibitem {Akerboom2015}S. Akerboom, J. Appel, D. Labonte, W. Federle, J.
Sprakel, \& M. Kamperman, (2015), Enhanced adhesion of bioinspired
nanopatterned elastomers via colloidal surface assembly. Journal of The Royal
Society Interface, 12(102), 20141061.

\bibitem {Kamperman}M. Kamperman, E. Kroner, A. del Campo, R.M. McMeeking, \& E. Arzt, (2010), Functional adhesive surfaces with ``gecko'' effect:
The concept of contact splitting. Advanced Engineering Materials, 12(5), 335-348.

\bibitem {Hwang}D. G. Hwang, K. Trent, \& M.D. Bartlett, (2018),
Kirigami-inspired Structures for Smart Adhesion. ACS applied materials \&
interfaces, 10(7), 6747-6754.

\bibitem {King}D.R. King, M.D. Bartlett, C.A. Gilman, D. J. Irschick,
\&  A. J. Crosby, (2014), Creating Gecko-Like Adhesives for ``Real World''
Surfaces. Advanced Materials, 26(25), 4345-4351.

\bibitem {Benz}M. Benz, K,J. Rosenberg, E. J. Kramer, \& J.N. Israelachvili, (2006). The deformation and adhesion of randomly rough and patterned
surfaces. The Journal of Physical Chemistry B, 110(24), 11884-11893.

\bibitem {Barreau}V. Barreau, R. Hensel, N.K. Guimard, A. Ghatak, R.M.
McMeeking, \& E. Arzt, (2016), Fibrillar elastomeric micropatterns
create tunable adhesion even to rough surfaces. Advanced Functional Materials,
26(26), 4687-4694.

\bibitem {Balijepalli}R. G. Balijepalli, M. R. Begley, N. A. Fleck, R. M. McMeeking, \& E. Arzt, (2016), Numerical simulation of the edge stress
singularity and the adhesion strength for compliant mushroom fibrils adhered
to rigid substrates. International Journal of Solids and Structures, 85, 160-171.

\bibitem {McMeeking}R. M. McMeeking, L. Ma, \& E. Arzt, (2010), Bi-Stable
Adhesion of a Surface with a Dimple. Advanced Engineering Materials, 12(5), 389-397.

\bibitem {PapCia2017}A. Papangelo, \&  M. Ciavarella, (2017). A
Maugis-Dugdale cohesive solution for adhesion of a surface with a dimple.
Journal of The Royal Society Interface, 14(127), 20160996.

\bibitem {PapCia2018}A. Papangelo, \& M. Ciavarella, (2018), Adhesion of
surfaces with wavy roughness and a shallow depression. Mechanics of Materials,
118, 11-16.

\bibitem {Labonte}D. Labonte, \& W. Federle, (2016), Biomechanics of
shear-sensitive adhesion in climbing animals: peeling, pre-tension and
sliding-induced changes in interface strength. Journal of The Royal Society
Interface, 13(122), 20160373.

\bibitem {Gravish}N. Gravish, M. Wilkinson, \& K. Autumn, (2008), Frictional
and elastic energy in gecko adhesive detachment. Journal of The Royal Society
Interface, 5(20), 339-348.

\bibitem {Savkoor}A. R. Savkoor, \& G. A. D. Briggs, (1977), The effect of
tangential force on the contact of elastic solids in adhesion. Proc. R. Soc.
Lond. A, 356(1684), 103-114.

\bibitem {Hutchinson1990}J. W. Hutchinson, (1990), Mixed mode fracture
mechanics of interfaces. Metal-ceramic interfaces, 295-306.

\bibitem {HutchinsonSuo}J. W. Hutchinson, \& Z. Suo, (1992). Mixed mode
cracking in layered materials. In Advances in applied mechanics, vol. 29 (eds
J. W. Hutchinson \& T. Y. Wu), pp. 63-191. Boston, MA: Academic Press.

\bibitem {Johnson1996} K. L. Johnson, (1996), Continuum mechanics modeling of
adhesion and friction. Langmuir, 12(19), 4510-4513.

\bibitem {Johnson1997}K. L. Johnson, (1997), Adhesion and friction between a
smooth elastic spherical asperity and a plane surface. In Proceedings of the
Royal Society of London A: Mathematical, Physical and Engineering Sciences
(Vol. 453, No. 1956, pp. 163-179). The Royal Society.

\bibitem {Waters}J. F. Waters, \& P. R. Guduru, (2010),
Mode-mixity-dependent adhesive contact of a sphere on a plane surface. In
Proceedings of the Royal Society of London A: Mathematical, Physical and
Engineering Sciences (Vol. 466, No. 2117, pp. 1303-1325). The Royal Society.

\bibitem {Schallamach}A. Schallamach, (1971), How does rubber slide?. Rubber
Chemistry and Technology, 44(5), 1147-1158.

\bibitem {Sahli}R. Sahli, G. Pallares, C. Ducottet, I.B. Ali, S. Al Akhrass, M. Guibert, \& J. Scheibert, (2018), Evolution of real contact area under
shear and the value of static friction of soft materials. Proceedings of the
National Academy of Sciences, 115(3), 471-476.

\bibitem {GJP}M. Ciavarella, \& A. Papangelo, (2018), A generalized Johnson
parameter for pull-off decay in the adhesion of rough surfaces. Physical
Mesomechanics, 21(1), 67-75.

\bibitem {Afferrante}M. Ciavarella, A. Papangelo, \& L. Afferrante, (2017),
Adhesion between self-affine rough surfaces: Possible large effects in small
deviations from the nominally Gaussian case. Tribology International, 109, 435-440.

\bibitem {violano}G. Violano, L. Afferrante, \& M. Ciavarella, (2018), On the
qualitative conflicts between stickiness criteria for hard multiscale randomly
rough surfaces, submitted.
\end{thebibliography}
\end{document}